\documentclass[final, 12pt,authoryear, nonatbib, times]{elsarticle}

\usepackage{amssymb}
\usepackage{amsmath}
\usepackage[font={normal}, justification=justified]{caption}
\usepackage[a4paper, total={6.5in, 9in}]{geometry}

\usepackage{longtable}
\usepackage{float}

\usepackage{url}
\journal{arXiv}

\begin{document}
% \linenumbers
\begin{frontmatter}

\title{Statistical analysis to assess porosity equivalence with uncertainty across additively manufactured parts for fatigue applications}

\author[c1]{Justin P. Miner}
\author[c1]{Sneha Prabha Narra\corref{cor1}}

\affiliation[c1]{organization={Mechanical Engineering Dept.},
            addressline={Carnegie Mellon University, 5000 Forbes Avenue}, 
            city={Pittsburgh},
            postcode={15213},
            state={PA},
            country={USA}}
            
\cortext[cor1]{Corresponding author. Email: snarra@andrew.cmu.edu}

\begin{abstract}
 
Previous work on fatigue prediction in Powder Bed Fusion - Laser Beam has shown that the estimate of the largest pore size within the stressed volume is correlated with the resulting fatigue behavior in porosity-driven failures. However, single value estimates for the largest pore size are insufficient to capture the experimentally observed scatter in fatigue properties. To address this gap, in this work, we incorporate uncertainty quantification into extreme value statistics to estimate the largest pore size distribution in a given volume of material by capturing uncertainty in the number of pores present and the upper tail parameters. We then applied this statistical framework to compare the porosity equivalence between two geometries: a 4-point bend fatigue specimen and an axial fatigue specimen in the gauge section. Both geometries were manufactured with the same process conditions using Ti-6Al-4V, followed by porosity characterization via X-ray Micro CT. The results show that the largest pore size distribution of the 4-point bend specimen is insufficient to accurately capture the largest pore size observed in the axial fatigue specimen, despite similar dimensions. Based on our findings, we provide insight into the design of witness coupons that exhibit part-to-coupon porosity equivalence for fatigue.

\end{abstract}

%%Research highlights
\begin{highlights}
\item Incorporated uncertainty in Generalized Pareto Distribution parameters. 
\item Applied to X-ray Micro CT data of Additively Manufactured fatigue specimens.
\item Estimated the largest pore size distribution with model uncertainty.
\item Largest pore size distribution varies with specimen geometry.
\end{highlights}

\begin{graphicalabstract}
\includegraphics[width=6.5in, height = 4.5in]{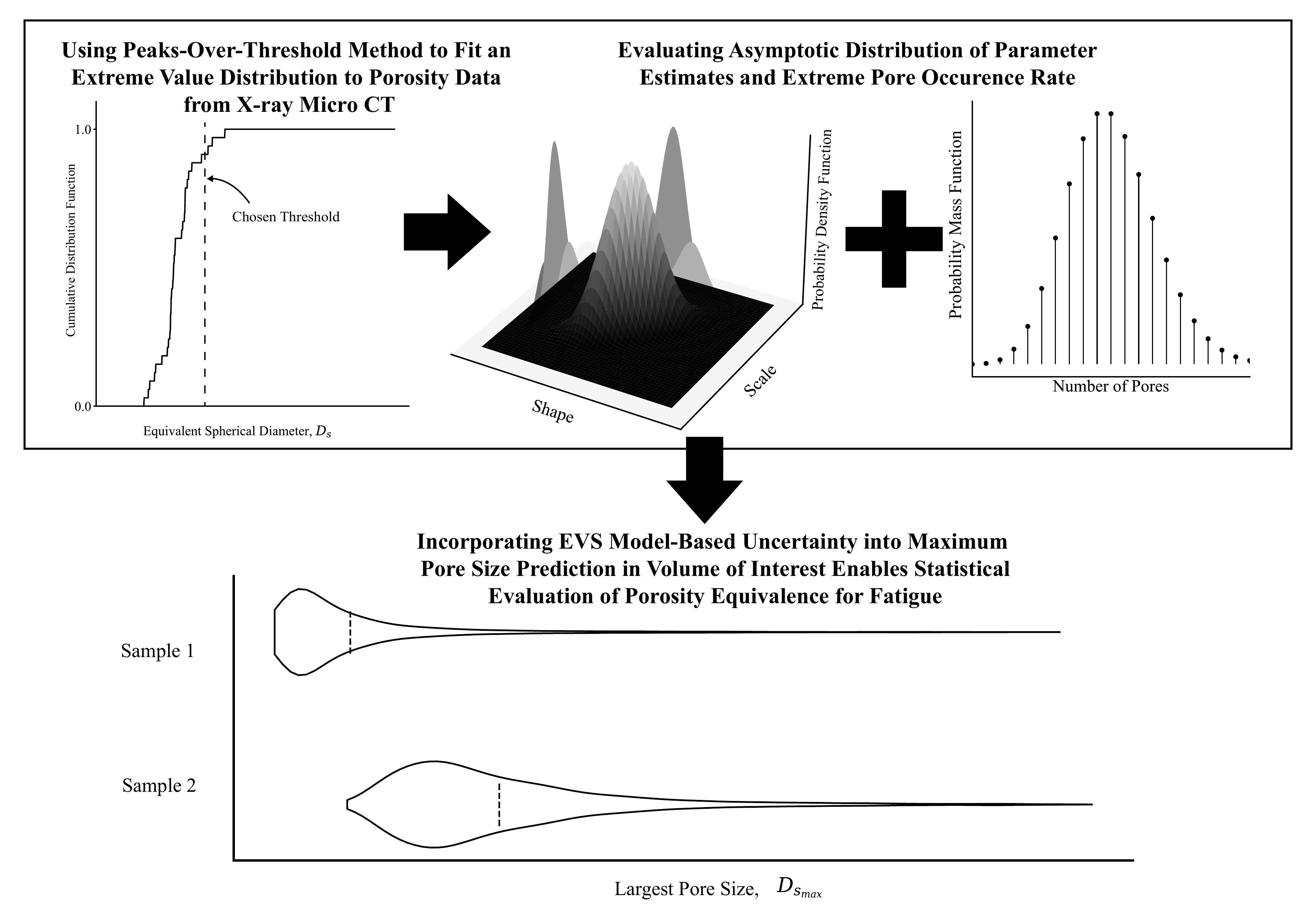}
\end{graphicalabstract}


\end{frontmatter}

\section{Introduction}

Powder Bed Fusion - Laser Beam (PBF-LB) is an additive manufacturing process that creates a near-net shape component by selectively melting thin layers of metal powder. Components fabricated with PBF-LB often contain defects, particularly porosity, which can affect the fatigue life. Porosity typically arises from mechanisms such as lack-of-fusion (LoF) \cite{Tang2017}, unstable keyholing \cite{Cunningham2017}, and powder-induced gas porosity \cite{Cunningham2017, Iebba2017}, among other unmodeled causes reported in \cite{Snow2023, MATTHEWS201633}.

In components experiencing fatigue loading, the critical pore size in the stressed volume has been correlated with the resulting fatigue behavior \cite{BEREZ2022106737, Macallister2022, PESSARD2021106206, LIU2022106700}. To model the size of critical defects, referred to as the largest pore size in this work, both extreme value statistics (EVS) \cite{BERETTA2021106407} and a number density approach \cite{REDDY2024108428} have been used in prior works. On the one hand, extreme value statistics approaches are generally applicable when the volume analyzed is less than or equal to the volume considered for prediction, where more than 1 pore in the upper tail is predicted. On the other hand, the number density approach is applicable when the volume scanned is greater than the volume considered for prediction, in cases where at least 1 pore per volume of interest is observed. For a witness coupon approach, generally a smaller or equivalent volume is analyzed to estimate the largest pore size in an actual part/component. Thus, in this work, we consider the EVS approach. In addition, the probability distribution parameters in the EVS approach allow for uncertainty quantification throughout the analysis.

Extreme value statistics was first applied to fatigue prediction in metallic inclusions in steel by Murakami in 1994 \cite{murakami1994inclusion}. This framework has since been applied to predict fatigue performance from porosity data in PBF-LB manufactured materials. \cite{BERETTA2021106407, TANG2019479, BERETTA2017178, ROMANO201732, ROMANO2018165, MINERVA2023112392}. There are two extreme value theorems. The first of these theorems is the Fisher-Tippett-Gnedenko  theorem which established that the maximum of a random variable in a given region follows the Generalized Extreme Value Distribution \cite{Fisher_Tippett_1928, gnedekno}, which is called the block maxima method. The second of these theorems is the Pickands-Balkema-De Haan theorem, which established that the maximum of a random variable above a threshold follows a Generalized Pareto Distribution \cite{pickands, balkemadehaan}, which is called the peaks-over-threshold (POT) method. In this study, the POT method was considered to capture the largest pore-generating phenomena and to prevent having to bin the data into "blocks" \cite{SHAHABI2022112027}, and it is easily formulated into a piecewise distribution because it captures the behavior above a chosen threshold. The GPD consists of multiple  parameters that describe the probability distribution \cite{EVSBook}. Using these parameters, a sensitivity analysis was performed and the predicted fatigue life was found to vary as a function of the upper tail parameters \cite{ANDERSON201878}. Furthermore, variability in fatigue properties from multiple initiation sites was found using a GPD approach combined with a damage-tolerant model. Specifically, the authors sampled two pores from an extreme value distribution and their locations from X-ray micro computed tomography (X-\textmu CT) data, and this was incorporated into a calibrated NASGRO model \cite{Macallister2022}. These studies demonstrate the importance of considering the largest pore size estimates beyond a single value. However, to date, the uncertainty in the upper tail parameters of an extreme value distribution model for porosity has not been incorporated into the prediction of the critical pore size and fatigue estimates. Thus, understanding and incorporating this uncertainty in the upper tail parameters is essential to capture any additional variation in fatigue properties that has not been accounted for with single value estimates such as the largest pore size \cite{BERETTA2017178} or the size metric from the number density approach \cite{REDDY2024108428}.

A case where uncertainty quantification in the largest pore size estimate is crucial is in understanding the correspondence in porosity between a witness coupon and a bulk part when evaluating fatigue properties. For instance, NASA Standard 6030 Section 4.11 recommends using witness coupons with the same geometry as the desired component to ensure equivalent material properties \cite{NASA6030}. Although feasible for small components, this approach is impractical for larger, more complex geometries or short-run, custom components, which are typical AM applications \cite{DEVSINGH2021350, SINGH20203058, LEARY2021597}. Additionally, a slight change in the largest pore size distribution in a given volume in different parts could produce significant differences in fatigue properties that are undetectable without extensive testing and characterization. Thus, analysis methods to capture the variability in the size of critical flaws have been identified as necessary for the probabilistic evaluation of fatigue in additively manufactured components \cite{Park2024}.

\begin{figure}[H]
    \centering
    \includegraphics[width=.75\textwidth]{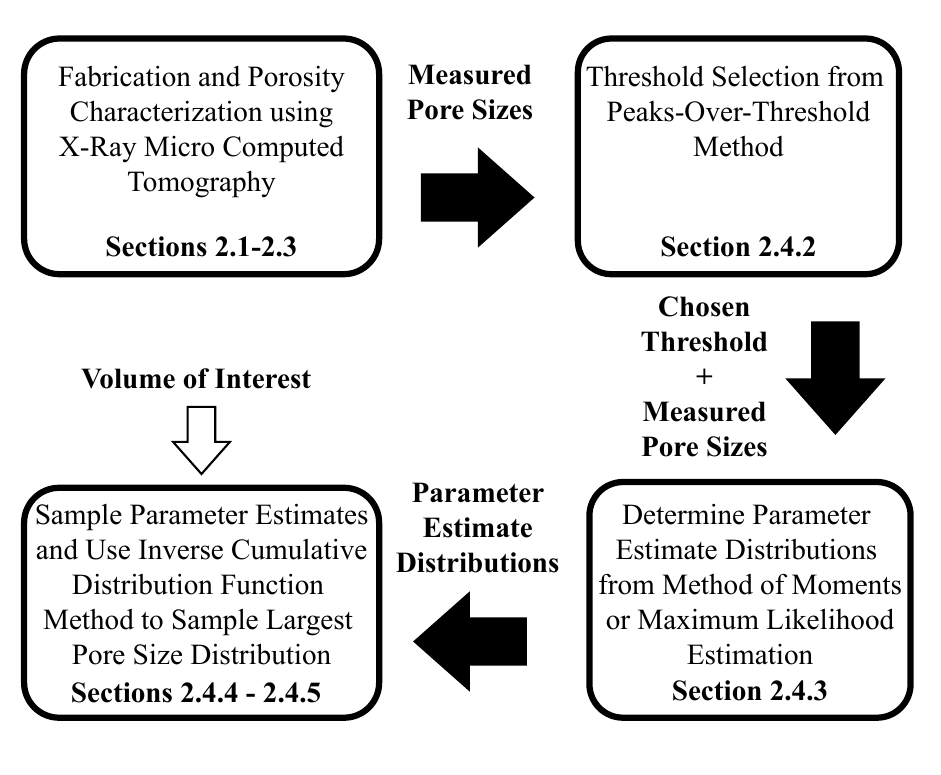}
    \caption{Flow chart showing the overall workflow and relevant sections. Note that the the code to utilize this workflow is made available upon publication.}
    \label{fig:method}
\end{figure}

In this work, we develop a method to incorporate model-based uncertainty into an EVS framework, enabling enhanced probabilistic estimation of the largest pore size present in a volume of interest. Figure \ref{fig:method} shows the workflow as well as highlights relevant sections of the manuscript. The workflow begins with the porosity data collection from 4-point bend fatigue specimens (4PB) using X-\textmu CT scanning. Then, EVS was used to fit the upper tail to the pore size distribution, followed by uncertainty quantification to incorporate variance in the porosity distribution parameters and number of pores that are expected top be present in a volume of interest, which is typically larger than the observed volume. By incorporating uncertainty in the distribution, we made comparisons between the largest pore size distributions estimated from the 4PB fatigue specimens against the largest pore sizes observed in the axial fatigue specimens.

\section{Materials and Methods}

\subsection{Powder Specifications}

The chemical composition of Ti-6Al-4V powder used in this work is shown in Table \ref{tab:Ti64}. The powder used has a nominal particle size distribution detailed in Table \ref{tab:powder}, with D\textsubscript{10}, D\textsubscript{50}, and D\textsubscript{90} indicating the 10\textsuperscript{th}, 50\textsuperscript{th}, and 90\textsuperscript{th} percentiles of particle sizes, respectively, which was verified using X-\textmu CT.

\begin{table}[H]
    \centering
    \caption{Chemical composition of Ti-6Al-4V (wt.\%). as provided by the powder manufacturer.}
    \label{tab:Ti64}
    \begin{tabular}{|c|c|c|c|c|c|c|c|}
        \hline
        Ti & Al & V & N & C & H & Fe & O \\
        \hline
         Balance & 6.09 & 3.89 & 0.013 & 0.011 & 0.0008 & 0.22 & 0.149  \\
         \hline
    \end{tabular}
\end{table}

\begin{table}[H]
    \centering
    \caption{Powder size distribution (\textmu m).}
    \label{tab:powder}
    \begin{tabular}{|c|c|c|}
        \hline
        D\textsubscript{10} & D\textsubscript{50} & D\textsubscript{90} \\
        \hline
         24 & 48 & 61\\
         \hline
    \end{tabular}
\end{table}

\subsection{Design of Experiments}

\begin{figure}[H]
    \centering
    \includegraphics[width=.85\textwidth]{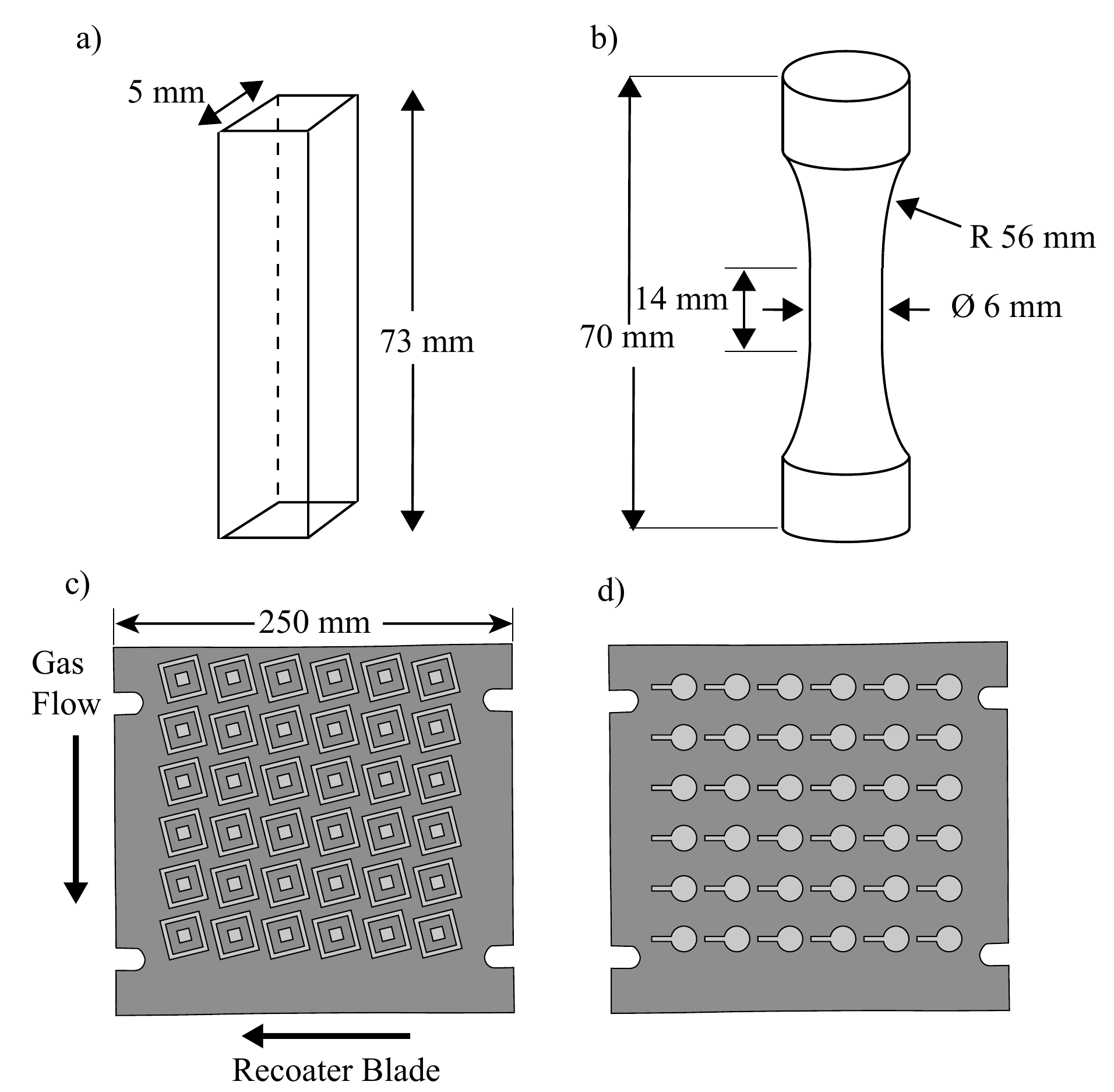}
    \caption{Drawings showing a) 4PB bend specimen b) axial specimen showing post machining and grinding dimensions. The layout of the specimens on the EOS M290 build plate build plate for c) 4PB specimen fabrication and d) axial specimen fabrication.}
    \label{fig:geo}
\end{figure}

The dataset utilized for this work consisted of X-\textmu CT scans from two different fatigue specimen geometries: a 4PB and an axial specimen as shown in Figs. \ref{fig:geo}a) and \ref{fig:geo}b). Fabrication was performed on an EOS M 290 PBF-LB machine. The constant process parameters used are included in Table \ref{tab:processparams}. A snake scan strategy was used for each layer. All fatigue specimens were printed with the longest dimension along the build direction for consistency. The only process parameter that varied was the scanning velocity. The scanning velocity ranged from 800 to 1900 mm/s in increments of 50 mm/s. The motivation for this was to systematically vary the source of porosity, the size and shape of the pores, and the total number of pores. The layouts of these geometries on the build plates are shown in Figs. \ref{fig:geo}c) and \ref{fig:geo}d).

\begin{table}[H]
    \centering
    \caption{PBF-LB process parameters used for fabrication of fatigue specimens in this work.}
    \label{tab:processparams}
    \begin{tabular}{|c|c|c|c|c|c|}
        \hline
        Laser Power & Hatch Spacing & Layer Height & Spot Size & Rotation Angle & Preheat\\
        (W)& (\textmu m)& (\textmu m) & (\textmu m) & (degrees) & (°C)\\
        \hline
         370 & 140 & 30 & 100 & 67 & 180\\
         \hline
    \end{tabular}
\end{table}

Given the shorter post-processing time for the 4PB fatigue specimens, these were manufactured with more combinations of velocities to explore the process space. Three builds of 36 4PB specimens were performed, while the axial specimens were manufactured with fewer combinations of velocities, 800, 1000, 1300, 1500, and 1800 mm/s, resulting in only one build of 36 specimens. After fabrication, the build plate containing each of these specimens was stress relieved according to ASTM Class B standards, which consisted of soaking the build plate for 2 $\pm$ 0.25 hours at 593°C $\pm$ 14°C followed by cooling in air \cite{ASTM2012, ams28012003heat}.

The 4PB specimens had dimensions of 6 x 6 x 73 mm\textsuperscript{3}. Surrounding each specimen, a thin wall, spaced 5 mm from the edge and 1 mm thick was placed. The as-built surface was machined to remove 500 \textmu m from each face and add a 500 \textmu m long 45° chamfer using low-stress grinding. This produced surface roughness $R_a < 0.2$ \textmu m. The dimensions of the finished specimen were 5 x 5 x 73 mm\textsuperscript{3} as shown in Fig. \ref{fig:geo}a). The axial specimens were manufactured according to ASTM E466-21 5.2.1.1 \cite{ASTM2002}. The as-fabricated specimen had a gauge diameter of 7 mm. Each specimen contained a sail to support the top half of the specimen. Similarly to the 4PB specimens, each specimen was finished by grinding 500 \textmu m off each surface by low stress grinding. This produced surface roughness $R_a < 0.2$ \textmu m. Each finished specimen had a gauge length of 14 mm, an inner diameter of 6 mm, and a fillet radius of 56 mm with an overall length of 70 mm as shown in Fig. \ref{fig:geo}b). 

\subsection{Porosity Characterization using X-\textmu CT}

Following post-processing, fatigue specimens were down-selected to be scanned using X-\textmu CT. For each geometry, the velocities and the number of specimens characterized for each geometry are shown in Table \ref{tab:couponcounts}. To allow equivalence to be evaluated for the most cases possible, this work used all X-\textmu CT data that was collected at these process parameters. Each specimen was scanned using a Zeiss Crystal CT, an X-\textmu CT device. The total height scanned for each specimen was greater than 5 mm with the total cross section captured for the 4PB specimen. For the axial specimens, a total volume greater than 200 mm\textsuperscript{3} was scanned from the gauge section with the total area of the gauge captured. These scans were cropped to 200 mm\textsuperscript{3} for analysis. The projections were captured at a source voltage of 150 kV and a power of 10 W. In total, for each scan 3001 projections consisting of 5 frame averages of 0.5 s exposure were captured. The HE2 source filter was used, along with a secondary reference taken with the HE18 filter to more effectively remove ring artifacts. The source-to-specimen distance was 19 mm and the specimen-to-detector distance was 565 mm, which led to a resolution of 2.5 \textmu m edge length per cubic voxel. The projections were reconstructed using Zeiss's reconstruction software using the auto center shift and a beam hardening coefficient of 0.25 which mitigated beam hardening artifacts for all geometries. The X-\textmu CT device and software used for acquisition and reconstruction are proprietary to the manufacturer. All relevant operating parameters are included for reproducibility using the same setup. 

\begin{table}[H]
    \centering
    \caption{Number of specimens scanned using X-\textmu CT at each velocity by geometry.}
    \label{tab:couponcounts}
    \begin{tabular}{|c|c|c|c|c|c|c|c|c|c|c|c|}
        \hline
        Velocity, $V$ (mm/s) & 800 & 850 & 900 & 950 & 1000 & 1050 & 1100 & 1150 & 1200 & 1250\\
        \hline
        Axial & 1 & 0 & 0 & 0 & 1 & 0 & 0 & 0 & 0 & 0\\
        \hline
        4PB & 1 & 1 & 1 & 1 & 1 & 1 & 1 & 1 & 2 & 1 \\
        \hline
    \end{tabular}

    \begin{tabular}{c}
    \\
    \end{tabular}
    
    \begin{tabular}{|c|c|c|c|c|c|c|c|c|c|c|}
        \hline
        Velocity, $V$ (mm/s) & 1300 & 1350 & 1400 & 1450 & 1500 & 1550 & 1600 & 1650 & 1700 & 1750 \\
        \hline
        Axial & 2 & 0 & 0 & 0 & 2 & 0 & 0 & 0 & 0 & 0 \\
        \hline
        4PB & 2 & 1 & 1 & 1 & 1 & 2 & 1 & 1 & 1 & 1 \\
        \hline
    \end{tabular}

    \begin{tabular}{c}
    \\
    \end{tabular}

    \begin{tabular}{|c|c|c|c|c|c|c|c|c|c|c|c|c|c|c|c|c|c|c|c|c|c||c|c|}
        \hline
        Velocity, $V$ (mm/s) & 1800 & 1850 & 1900 \\
        \hline
        Axial & 2 & 0 & 0 \\
        \hline
        4PB & 1 & 1 & 1 \\
        \hline
    \end{tabular}
\end{table}

To segment the porosity for each scan, a 3-dimensional (3D) U-Net was applied using Dragonfly Pro \cite{ObjectResearchSystemsORSInc2022}. This model was trained in previous work and is evaluated on a 64 x 64 x 3 voxel volume \cite{REDDY2024108428}. This method was used to segment the X-\textmu CT scans for increased repeatability and consistency compared to more traditional threshold-based methods \cite{unetct, Desrosiers2024}. Each connected component was separated and the volume, aspect ratio, and surface area were extracted. The surface area was computed through a summation of voxels. The equivalent spherical diameter was calculated from the volume using the following equation.

\begin{equation}
    D_s = (6V/\pi)^{1/3},
\end{equation}

\noindent where $D_s$ is the equivalent spherical diameter and $V$ is the volume determined by counting the number of voxels contained within a pore. The equivalent spherical diameter is the diameter of a sphere with the same volume as a given pore. 

The aspect ratio can be calculated using the following equation.

\begin{equation}
    AR = a/b
\end{equation}

\noindent where $AR$ is the aspect ratio, $a$ is the smallest dimension or minimum Feret diameter, and $b$ is the largest dimension or maximum Feret diameter. The sphericity, $\psi$, was calculated using the following equation.

\begin{equation}
    \psi = \pi^{1/3}(6V)^{2/3}/A,
\end{equation}

\noindent where $A$ is the surface area. The sphericity is a measure of how closely a pore is to a sphere according to the surface area and volume ratio. A perfect sphere will have a sphericity of 1. The further a pore is from a spherical morphology, the smaller the value of sphericity.

\subsection{Upper Tail Distribution Characterization}

\subsubsection{Assumptions}

\noindent The following simplifying assumptions are made to allow for modeling.

\begin{itemize}
    \item EVS model-based uncertainty was the only source of uncertainty considered.
    \item The size of the pores can be represented as an independent and identically distributed (I.I.D.) random variable.
    \item There is no error in the chosen threshold from the EVS fit.
\end{itemize}

From the assumption that pore sizes are I.I.D., the occurrence of pores with size above the threshold follows a stationary Poisson process. This has been carried out successfully in other EVS literature on extreme weather events \cite{Poisson, rosbjerg1992prediction, HU2023108935}. Thus, the number of pores in a larger volume can be modeled as a Poisson distribution with constant rate, $\lambda$. This rate can be estimated using the number of pores above the threshold divided by the volume. The variance of this parameter estimate is $\lambda/n$, where $n$ is the number of pores that were captured above the threshold.

\subsubsection{Threshold Selection}
\label{sec:ts}

To determine the region of stability in the threshold, the mean excess function is used to determine a range of thresholds at which the bias-variance trade-off is optimal. The mean excess function is defined as the mean of all data points above a chosen threshold. The theoretical mean excess function for the GPD can be expressed as, 
\begin{equation}
    M(\mu) = \sigma/(1-\xi) + \xi/(1-\xi) \mu,
\end{equation}

\noindent where $M$ is the mean excess function, $\mu$ is the threshold, $\sigma$ is the scale parameter, and $\xi$ is the shape parameter. This function is only defined for $\xi \le 1$. Otherwise, the mean of the GPD that is fit would be infinite, and a mean excess function would not be applicable \cite{GHOSH20101492}. In the present study, for all the fatigue specimens analyzed, $\xi \le 1$, which makes physical sense because the mean porosity size within a given specimen cannot be infinite. 

To determine the region of stability, the threshold needs to be large enough so a linear relationship is present in the mean excess function with threshold, while small enough to minimize variance by ensuring sufficient data points are used for fitting. This linear region can be observed in Fig. \ref{fig:method}a) along the straight line, marked in red. After estimating a region of threshold value (dashed line marked in black in Fig. \ref{fig:method}a), the stability in shape and modified scale parameters were evaluated in the region of interest of the threshold. The modified scale is derived from the asymptotic behavior of the scale as a function of the threshold as follows.

\begin{equation}
    \sigma^* = \sigma(\mu) - \xi \mu,
\end{equation}

\noindent where $\sigma(\mu)$ is the scale parameter as a function of the threshold, $\xi$ is the shape parameter, and $\sigma^*$ is the modified scale parameter \cite{EVSBook}. A smaller region of threshold selection where both the shape and modified scale remain approximately the same value with changing threshold is where the threshold should be chosen. This ensures that slight changes to the threshold would not affect the upper tail parameters, thereby improving the validity of our assumptions. Example plots showing the goodness of fit are included in Section \ref{sec:params}.

\subsubsection{Maximum Likelihood and Method of Moments Estimators}
\label{sec:mlemom}

Typically, the Maximum Likelihood Estimate (MLE) is used to estimate the shape and scale parameters of the GPD distribution after choosing the threshold. The MLE estimates the probability distribution parameter values that maximize the probability of observing the actual data under the assumed model, which is GPD in our problem. The probability distribution of the parameter estimates is as follows:

\begin{equation}
    \label{eq:covmle}
    \begin{bmatrix}\hat{\sigma}_{MLE} \\ \hat{\xi}_{MLE}  
    \end{bmatrix} \to \mathcal{N}\left(
    \begin{bmatrix}\sigma \\ \xi 
    \end{bmatrix}, \frac{1+\xi}{n}\begin{bmatrix}
        2\sigma^2 & \sigma \\
        \sigma & (1+\xi)
    \end{bmatrix}
    \right) \text{ as } n \to \infty,
\end{equation}

where $\mathcal{N}$ denotes the normal distribution, $\hat{\sigma}$ is the scale parameter estimate with a subscript that denotes the method used to estimate it being the MLE, $\hat{\xi}$ is the shape parameter estimate with the subscript denoting the method used to estimate it being the MLE, and n is the number of data points above the threshold. This estimator is found to be asymptotically normal when $\xi > -0.5$ \cite{BRAZAUSKAS2009424}. That is, with infinitely many points in the upper tail, the GPD distribution parameter estimates follow a bivariate normal distribution with the mean equal to the true value and the covariance matrix as specified when $\xi > -0.5$.

The Method of Moments (MOM) estimator can alternatively be used to estimate the shape and scale parameters of the GPD distribution from the choice of threshold. The method of moments fits the theoretical moments of the GPD which is a function of the parameters to the observed moments of the data. Since we are estimating two parameters, the first and second theoretical moments, which are the mean and variance, are used for estimation. The MOM estimator is asymptotically normal when $\xi < 0.25$ \cite{BRAZAUSKAS2009424}. The probability distribution of the parameters estimates is as follows:

\begin{equation}
    \label{eq:covmom}
     \begin{bmatrix}\hat{\sigma}_{MOM} \\ \hat{\xi}_{MOM}  
    \end{bmatrix} \to \mathcal{N}\left(
    \begin{bmatrix}\sigma\\ \xi  
    \end{bmatrix}, \frac{(1-\xi)^2}{n(1-3\xi)(1-4\xi)}\begin{bmatrix}
        2\sigma^2\frac{(1-6\xi + 12\xi^2)}{(1-2\xi)} & \sigma(1-4\xi + 12\xi^2) \\
        \sigma(1-4\xi + 12\xi^2) & (1-2\xi)(1-\xi+6\xi^2)
    \end{bmatrix}
    \right) \text{ as } n \to \infty.
\end{equation}

\noindent The MOM estimator is found to converge slower than the MLE estimator, however, the MOM is asymptotically normal in cases where $\xi \le -0.5$ where the MLE estimator is not \cite{BRAZAUSKAS2009424}. Thus, to estimate uncertainty in the parameters, in this work, both estimators were used within their proper domains (where they are estimated to be asymptotically normal).

\subsubsection{Modeling the Largest Pore Size as a Function of Random Variables}

The largest pore size distribution with a given number of pores above the threshold is as follows:

\begin{equation}
    P[D_{s_{max}} \le d] = F(d) = (1-(1+\xi\frac{d-\mu}{\sigma})^{-1/\xi})^N,
\end{equation}

\noindent where $D_{s_{max}}$ is the largest equivalent spherical diameter of a pore given that there are $N$ pores above threshold. The value of $d$ is the sample value that is well defined for $d > \mu$ when $\xi \ge 0$ and $\mu \le d \le \mu - \sigma/\xi$ for $\xi \le 0$ \cite{ROMANO201732}. 

The inverse cumulative distribution function (CDF) gives the $100 \cdot p$\textsuperscript{th} percentile pore size and can be calculated as follows:

\begin{equation}
    D_{s_{max}}(p) = \sigma/\xi \cdot ((1- p^{1/N})^{-\xi} -1)+\mu.
\end{equation}

\noindent where $N$ is the number of pores above the threshold. Given that the shape and scale are asymptotically normal according to the appropriate estimators, the number of pores above threshold can be represented as a Poisson distribution with a Gaussian estimate of rate, and, using the inverse CDF method, $p$ can be sampled as a standard uniform random variable, the largest pores can be modeled as a function of random variables, $\sigma, \xi, N, \text{ and } p$. Thus, $D_{s_{max}}$ can be sampled by sampling all of the aforementioned random variables given that $N > 0$ (more than one pore is present in the upper tail).

In the case where 0 pores above the threshold were sampled ($N = 0$), the empirical CDF was used to obtain $D_{s_{max}}$ by raising it to the power of the number of pores below the threshold (as a Poisson random variable) which follows from N\textsuperscript{th} order statistics \cite{nthorder}. This allowed for prediction even in low volumes of interest, removing this limitation of EVS.

\subsubsection{Monte Carlo Method}
\label{sec:montecarlo}

Because no closed-form solution from the integration of cumulative distribution functions can be calculated to determine the marginal cumulative density function of $D_{s_{max}}$, a Monte Carlo method was used to sample this distribution from the parameter estimate distributions. Similar methods have been published in EVS literature to quantify uncertainty in the parameter estimates and its impact on the estimation of the largest occurrence in a given domain such as the largest potential loss in stock prices within a time interval \cite{Huang2012}, the largest river flow rate for a fixed time period \cite{das2016characterization}, and the 50-year wind speeds \cite{evsmc}. In this work, to sample the largest pore size distribution in a volume of interest, 1000 $N$ values, 1000 $p$ values, and 1000 $(\sigma, \xi)$ pairs were sampled from the previously specified distributions. Following this, the $D_{s_{max}}$ values were calculated for each combination of points. Then, the probability density function was computed using a histogram approximation. Finally, the cumulative density function was computed through numerical integration. In total, 1 billion samples were used to determine the CDF of $D_{s_{max}}$ for the 4PB samples considering uncertainty in parameter estimates. This code was optimized using CuPy, which is a gpu-optimized adaptation of NumPy and SciPy \cite{cupy_learningsys2017}. On an RTX490 GPU, this code took approximately five minutes (in clock time) per fatigue specimen to execute and used 9 GB of GPU memory.

\section{Results}

\subsection{Poisson and GPD Parameter Estimates}
\label{sec:params}

\begin{figure}[H]
    \centering
    \includegraphics[width=0.9\textwidth]{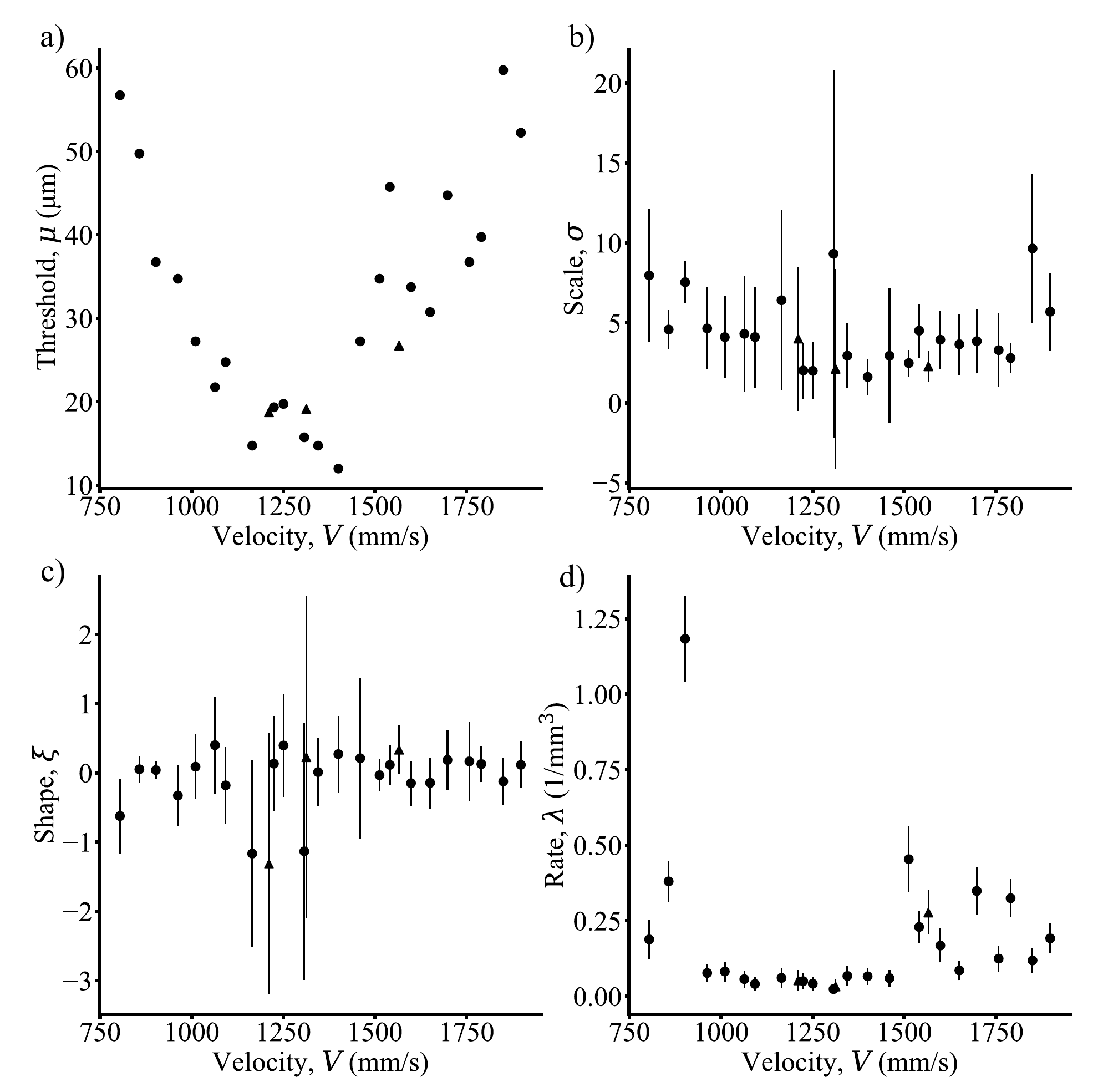}
    \caption{Parameter estimates for a) threshold, b) scale, c) shape, and d) pore rate along with a 95\% confidence interval for these estimates shown for b) scale, c) shape, and d) rate. At 1300 and 1550 mm/s, the estimated parameters from different specimens with the same process parameters are denoted by different shaped markers.}
    \label{fig:paramests}
\end{figure}

Figure \ref{fig:paramests} shows the parameter estimates by velocity for the 4PB specimens, with different specimens processed with the same conditions as denoted by differently shaped markers. Figure \ref{fig:paramests}a) shows the chosen threshold. It can be seen that the estimated threshold is in correspondence with the process regime. That is, the threshold increases at velocities below 1100 mm/s and above 1400 mm/s which correspond to the occurrence of keyhole and LoF porosity, respectively. Figures \ref{fig:paramests}b) and \ref{fig:paramests}c) show the estimated scale and shape parameter estimates for the threshold selection in Fig. \ref{fig:paramests}a). It should be noted that the error bars in these figures represent the marginal 95\% confidence interval of the respective parameter estimates. The word marginal is used because there is a non-zero covariance between the shape and scale from Equations \ref{eq:covmle} and \ref{eq:covmom}, indicating that the shape and scale parameters are correlated and do not vary independently of each other, as is seemingly implied by the graph. Lastly, Fig. \ref{fig:paramests}d) shows the pore rate estimate, which gives the total number of pores above the threshold divided by the volume. The confidence interval for this estimate comes from the variance according to the MLE of the Poisson rate parameter. Overall, the uncertainties present in the shape, scale, and pore occurrence rate can be propagated into the largest pore size expected in a given volume, leading to significant uncertainty.

\begin{figure}[H]
    \centering
    \includegraphics[width=0.9\textwidth]{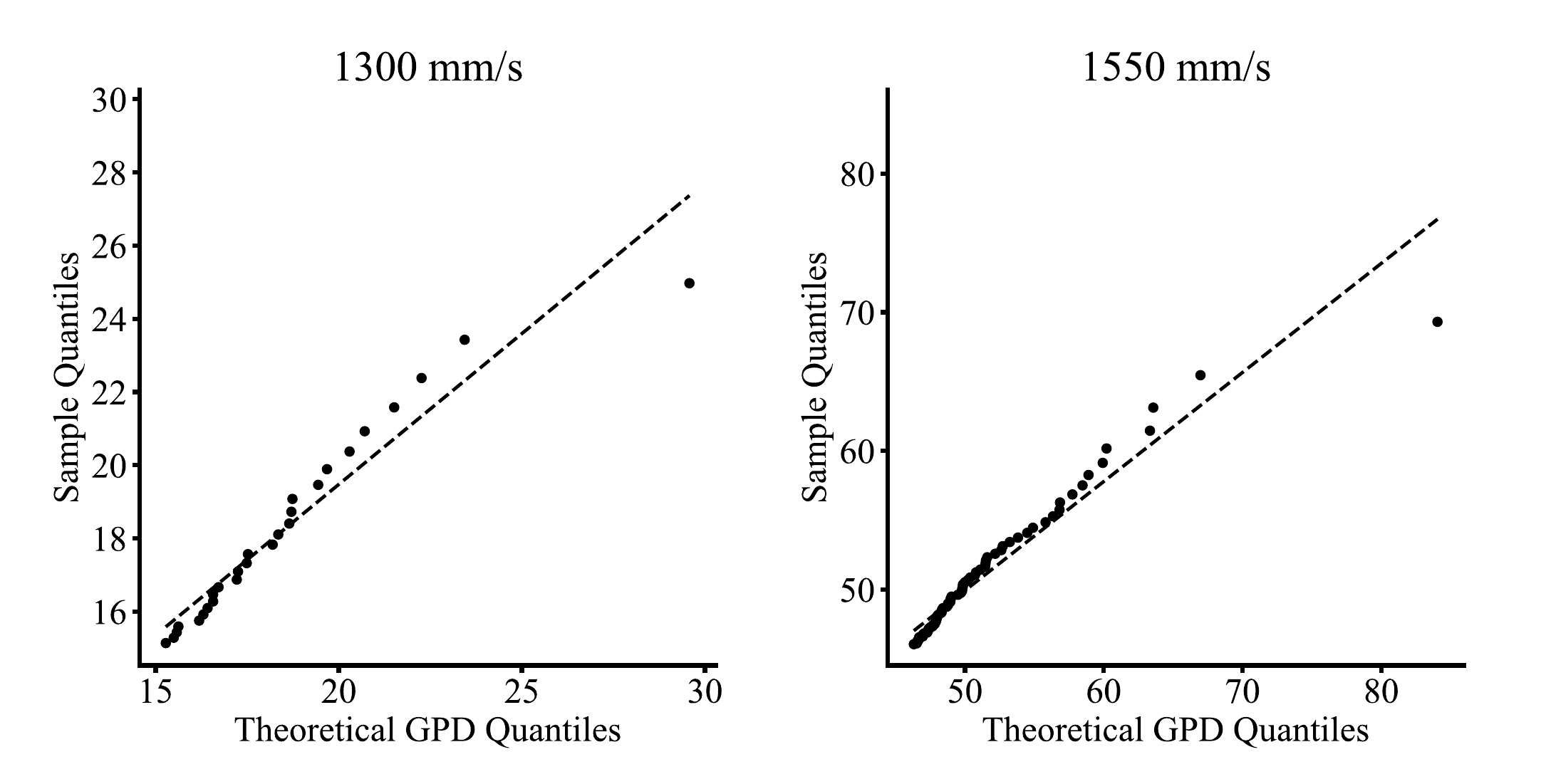}
    \caption{Quantile-quantile plots showing the quality of fit of the data (sample quantiles) to the theoretical GPD (theoretical quantiles) for a sample at 1300 and 1550 mm/s.}
    \label{fig:qq}
\end{figure}

In some specimens, large confidence intervals in the scale and shape parameter estimates are present despite containing a sufficient number of points present in the upper tail. Figure \ref{fig:qq} shows the Quantile-Quantile (QQ) plots at 1300 mm/s, where large error bars in shape and scale are observed and at 1550 mm/s where small error bars are observed in these parameter estimates. QQ plots allow for assessment of the quality of fit of a theoretical GPD (theoretical quantiles) against the observed pore sizes (sample quantiles). In each plot, a good quality of fit is present. More than 30 points are in the upper tail which indicates that sufficient points are present. Nonetheless, the uncertainty present in the shape and scale reflects the porosity data present at these points and is necessary in order to capture the upper tail behavior that could result from the data.

\subsection{Evaluating Pore Size Equivalence Between Geometries}

\begin{figure}[H]
    \centering
    \includegraphics[width=1\textwidth]{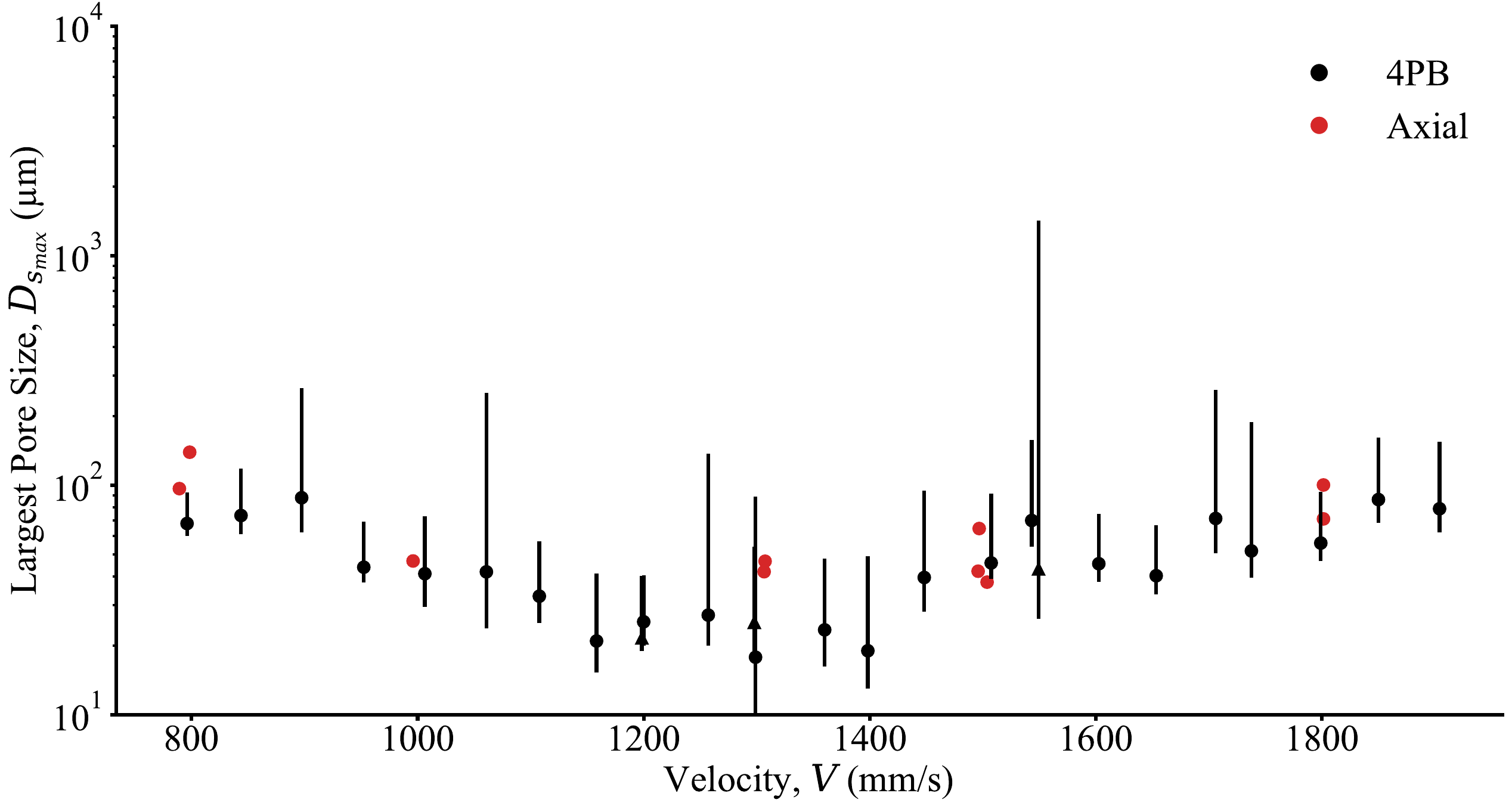}
    \caption{Estimated largest pore size distribution using porosity data from each 4PB specimen. At 1300 and 1550 mm/s the marker shape differentiates between different specimens at the same process parameters which corresponds to Fig. \ref{fig:paramests}. The black point represents the mean predicted size and the bands reflect the 95\% confidence interval. The red points represent the largest pore size observed in axial specimens with the volume scanned (200 mm\textsuperscript{3}).}
    \label{fig:sizepreds}
\end{figure}

This section discusses the application of the EVS+UQ framework to evaluate part-to-coupon equivalence for porosity-driven fatigue properties. For this purpose, we will consider the 4PB geometry as the "coupon" and the axial geometry as the "part". Figure \ref{fig:sizepreds} shows the largest pore size distribution estimated from the porosity data in the 4PB geometry for a 200 mm\textsuperscript{3} volume of interest, to represent the scanned region of the gauge section in the axial specimens. The same figure also shows the largest pore size observed in the axial specimens from the X-\textmu CT data. The 95\% confidence interval of the largest estimated pore size was determined by the Monte Carlo method (described in Section \ref{sec:montecarlo}). A difference in the estimated largest pore size distribution and the observed largest pore size can be seen at 800, 1300, and 1800 mm/s. Specifically, the largest observed pore size in the axial specimens is relatively higher than the estimated largest pore size. If the porosity distribution was the same between the two fatigue specimen geometries, there would be minimal difference between the estimated largest pore size distribution and the observed largest pore size and the observed value would consistently fall within the estimated confidence interval. This result has significant implications in the design of an acceptable witness coupon to achieve a part-to-coupon porosity equivalence for fatigue, specifically in the qualification of PBF-LB components using smaller witness coupons.

\section{Discussion}

\subsection{Individual Contributions of Uncertainty Sources}

\begin{figure}[H]
    \centering
    \includegraphics[width=1\textwidth]{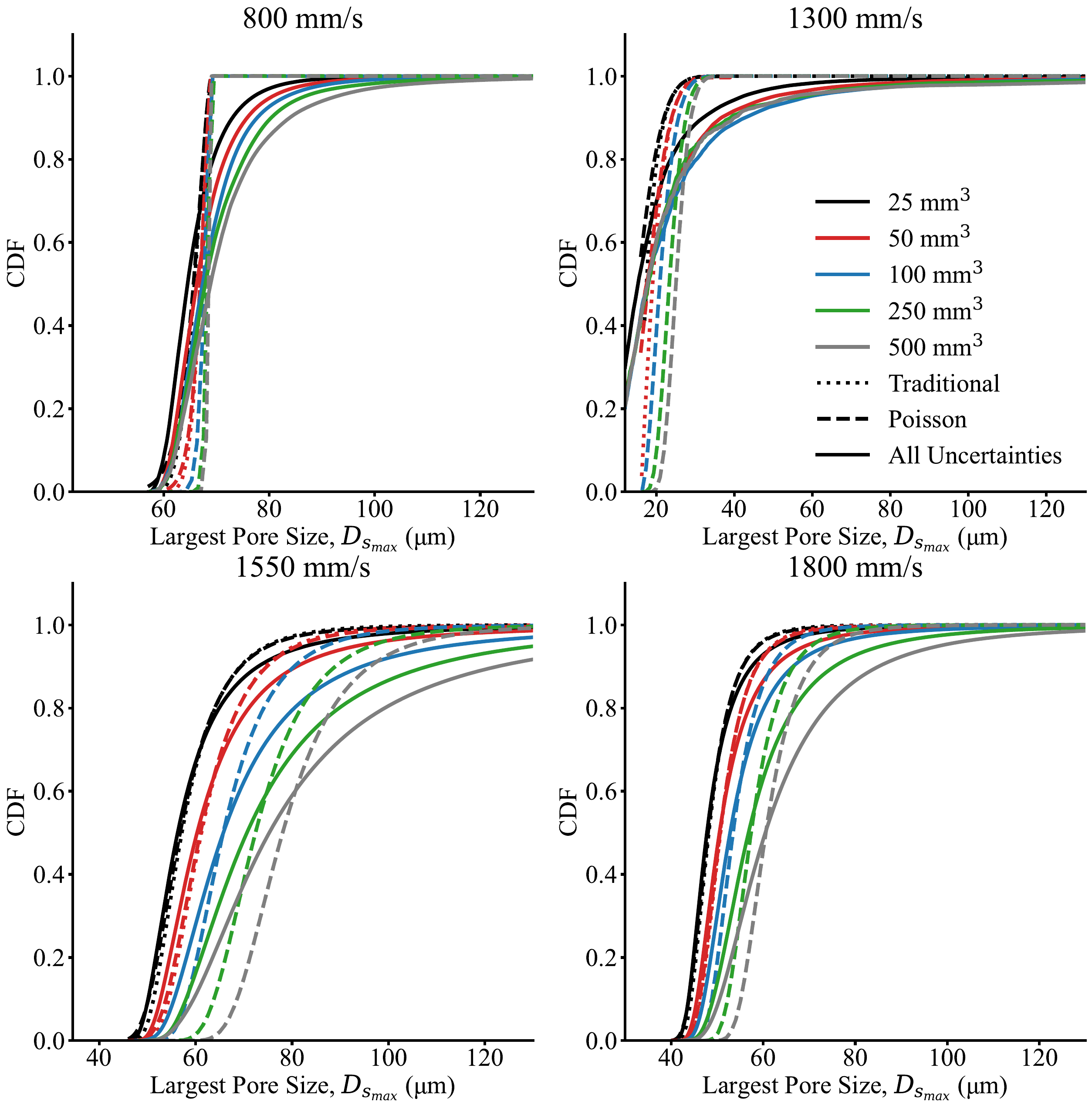}
    \caption{CDFs of the largest pore size distribution at different process parameters for different volumes indicated by color and by propagating different sources of uncertainty indicated line type.}
    \label{fig:uq}
\end{figure}

To understand how individual elements of uncertainty impact the largest pore size distribution in different volumes of interest, Fig. \ref{fig:uq} shows the CDF of the largest pore size in 4PB fatigue specimens fabricated at scanning velocities of 800, 1300, 1550, and 1800 mm/s. Each subplot shows the traditional CDF of the largest pore size (using the point estimate of parameters), the CDF that incorporates the uncertainty in the pore number (denoted as Poisson), and the CDF that incorporates the uncertainty in the upper tail parameters and pore number (denoted as All Uncertainties) for different extrapolated volumes. 

A distinction should be made for volumes less than 100 mm\textsuperscript{3} and volumes greater than 100 mm\textsuperscript{3}. In the smaller volumes shown, the volume of interest is less than that scanned with X-\textmu CT. Where in larger volumes, the volume of interest is greater than what was scanned using X-\textmu CT. The latter is what has been traditionally incorporated in materials science datasets because generally it is desired to predict behavior at larger volumes. However, the former (predicting within smaller volumes than scanned) is a valid approach for using the GPD as long as one or more extreme events (pores in this case) are expected within the desired interval (volume in this case). Such predictions within a smaller interval have been carried out in environmental research \cite{smallinterval}.

In Fig. \ref{fig:uq}, as the volume of interest decreases, the largest pore size decreases, as expected, and the maximum absolute difference (or Kolmogorov–Smirnov (KS) statistic) between the traditional CDF (without uncertainty) and the CDF considering the uncertainty in the pore number (Poisson) increases. This change in distance between CDFs with different sources of uncertainty is quantified in Table \ref{tab:ks}. It can be seen that at all scanning velocities except 800 mm/s, as the volume of interest increases, the KS statistic between the traditional CDF and the CDF considering uncertainty in pore number (Poisson) decreases while the KS statistic between the traditional CDF and the CDF including all sources of uncertainty considered increases. At 800 mm/s, with increasing volume of interest, the KS statistic between the traditional CDF and the CDF considering uncertainty in pore number (Poisson) remains approximately the same while the KS statistic between the traditional CDF and the CDF including all sources of uncertainty considered increases. This indicates that with increasing volume of interest, the uncertainty due to the parameter estimates of shape and scale are more significant than the uncertainty due to the number of pores.

\begin{table}[H]
    \caption{Table of KS statistic between each CDF with uncertainty and the traditional CDF at the volumes and scanning velocities shown in Fig. \ref{fig:uq}. The KS statistic between the CDF with all sources of uncertainty and the traditional CDF is denoted by the column "all".}
    \label{tab:ks}
    \centering
    \begin{tabular}{|c|c|c|c|c|c|c|c|c|}
        \hline
        Volume (mm\textsuperscript{3}) & \multicolumn{8}{c|}{Scanning Velocity (mm/s)}\\
        \hline
         & \multicolumn{2}{c|}{800} & \multicolumn{2}{c|}{1300} & \multicolumn{2}{c|}{1550} & \multicolumn{2}{c|}{1800}\\
        \hline
        & Poisson & All & Poisson & All & Poisson & All & Poisson & All \\
        \hline
        25 & 0.06 & 0.22 & 0.23 & 0.41 & 0.05 & 0.08 & 0.04 & 0.06\\ 
        \hline
        50 & 0.04 & 0.28 & 0.32 & 0.40 & 0.03 & 0.10 & 0.01 & 0.07\\ 
        \hline
        100 & 0.08 & 0.38 & 0.17 & 0.45 & 0.02 & 0.19 & 0.01 & 0.12\\
        \hline
        250 & 0.16 & 0.40 & 0.09 & 0.52 & 0.01 & 0.19 & 0.01 & 0.16\\
        \hline
        500 & 0.09 & 0.49 & 0.06 & 0.61& 0.01 & 0.24 & 0.01 & 0.19\\
        \hline
    
    \end{tabular}
\end{table}

In larger volumes, the uncertainty in the largest pore size estimation is driven by uncertainty in the shape parameter of the fitted distribution. When the confidence interval of the shape parameter crosses zero, as is seen in most of the upper tails, it indicates that the largest pore size distribution could be heavy-tailed or weak-tailed, which introduces significant uncertainty as a result of the unpredictable nature of the tail. Thus, in order to effectively compare the largest pore size distribution, considering all sources of uncertainty in the modeling is necessary.

The method developed in this study considers various sources of uncertainty that would not be captured when considering only the largest pore size in a region. Using EVS, we quantified the largest pore size distribution that could occur because of the sampling of different regions of the same volume. Additionally, by sampling different regions of the same volume, the Poisson distribution is able to capture the distribution in the total number of pores, which governs how many pores could be found in the upper tail. Furthermore, by propagating the uncertainty in the parameter estimates, we are able to capture the different tails that could result. Thus, this method captures many sources of variability in the largest pore size.

\subsection{Difference between Largest Pore Size Distributions}

To compare the largest pore size observations in the axial specimens and the distributions estimated for the 4PB specimens, a $p$-value and $q$-value were computed as shown in Table \ref{tab:pvals}. The $p$-value is the probability that the largest pore size in a 4PB specimen would be farther from the mean as was observed for the axial specimen in the volume of interest (200 mm\textsuperscript{3}). The $q$-value is the quantile of the CDF of the largest pore size in which the largest axial pore lies. If the axial and 4PB specimens had the same largest pore size, most of the $p$-values would be closer to 1. If the largest pore size between samples distribution was the same, a standard uniform distribution would be expected in the $q$-value because the observations would effectively be sampling from the largest pore size distribution. Differences in distribution, thus, would be indicated by larger $p$-values and $q$-values that consistently deviate from 0.5.

The largest $p$-values and $q$-values closest to the median of 0.5 are at 1000 mm/s indicating good agreement between the point estimate from the 4PB specimen and the observation in the axial specimen. At 1500 mm/s, the $p$-values are low but the $q$-values vary, which corresponds to the large scatter in the largest pore sizes from the axial specimen in comparison to the estimated distribution from the 4PB specimen and more extreme pore sizes than estimated. At 800, 1300, and 1800 mm/s, small $p$-values are observed with large $q$-values indicating that the observations in the axial specimen are significantly larger than the estimated distribution in the 4PB specimen. At 800 and 1800 mm/s, this corresponds to an increase in keyhole pore size and lack-of-fusion pore size, respectively; however, at 1300 mm/s, where gas porosity is the only type of pore expected to be present, the increase in pore size warrants further investigation.

\begin{table}[H]
    \centering
    \caption{$p$-values that denote the probability that largest pore size observed (from axial geometry) has a larger difference from the mean of the estimated largest pore size distribution (from 4PB) than what is present. $q$-values denoting the quantile of the estimated distribution that the largest pore size observed lies in.}
    \label{tab:pvals}
    \begin{tabular}{|c|c|c|c|}
        \hline
        Scanning Velocity, $V$ (mm/s) & $p$-values & $q$-values\\
        \hline
         800 & 0.065, 0.006 & 0.967, 0.997\\
         \hline
         1000 & 0.893 & 0.554\\
         \hline
         1300 & 0.186, 0.147, 0.499, 0.439 & 0.907, 0.927, 0.751, 0.781\\
         \hline
         1500 & 0.132, 0.006, 0.195 & 0.934, 0.003, 0.098\\
         \hline
         1800 & 0.369, 0.063 & 0.815, 0.968\\
         \hline
    \end{tabular}
\end{table}

\subsection{Effect of Specimen Location on the Build Plate on Largest Pore Size Distribution}

\begin{figure}[H]
    \centering    
    \includegraphics[width=0.9\textwidth]{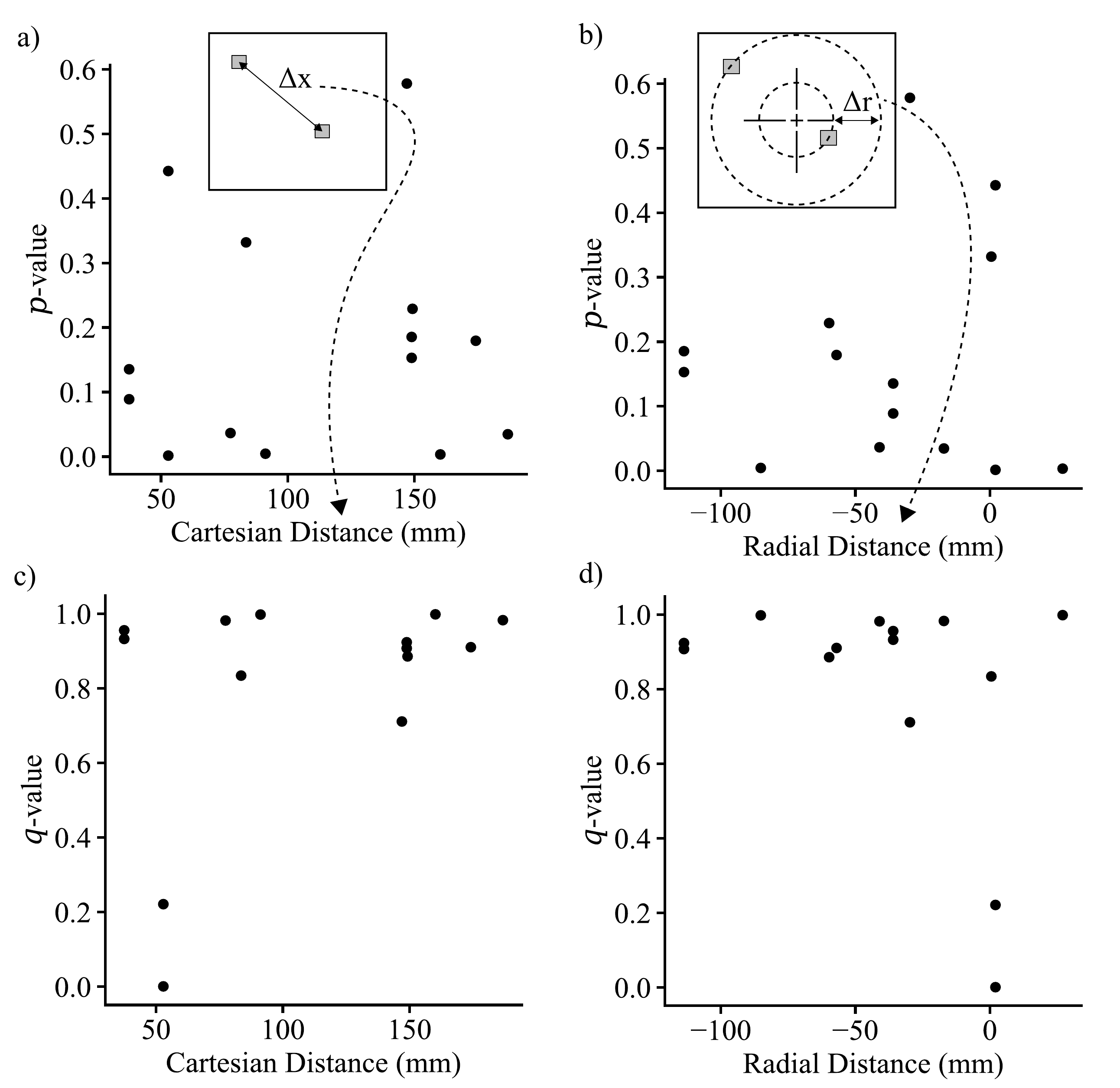}
    \caption{Scatter plots capturing the similarity in distribution as a function of the build plate location. The $p$-value related to the observation in the axial sample's distance from the mean of the 4PB test, and the $q$-value is the quantile of the CDF that the observation lies in. The two distance metrics are visually described in the figure.}
    \label{fig:pvals}
\end{figure}

Figure \ref{fig:pvals} compares the difference in location on the build plate among samples at the same process parameters with different geometry to determine whether the similarity in distribution was related to the location on the build plate. To determine the differences between 4PB and axial largest pore size distributions, two metrics are used, the $p$-value and $q$-value. These metrics are plotted against two distance measures. The Cartesian distance, shown in Figs. \ref{fig:pvals}a) and \ref{fig:pvals}c), was used to determine whether differences in the shielding gas flow had any effect on the largest pore size distribution. In the EOS M290, it was found that the distance from the gas flow inlet had an effect on the resulting porosity \cite{MORAN2021102333}, which would be captured by the Cartesian distance. The radial distance, as shown in Figs \ref{fig:pvals}b) and \ref{fig:pvals}d), was used to determine whether differences in the laser interaction angle had any effect. In all the subfigures in Fig. \ref{fig:pvals}, there is no clear trend observed. Thus, it was concluded that location was not found to have a significant impact on the differences in distribution observed within this dataset, and the differences in pore size observed at 1300 mm/s, where only gas porosity is expected, cannot be explained solely by differences in the location of the specimen on the build plate.

\subsection{Porosity Differences between Geometries}
\label{sec:largerpores}

\begin{figure}[H]
    \centering
    \includegraphics[width=0.9\textwidth]{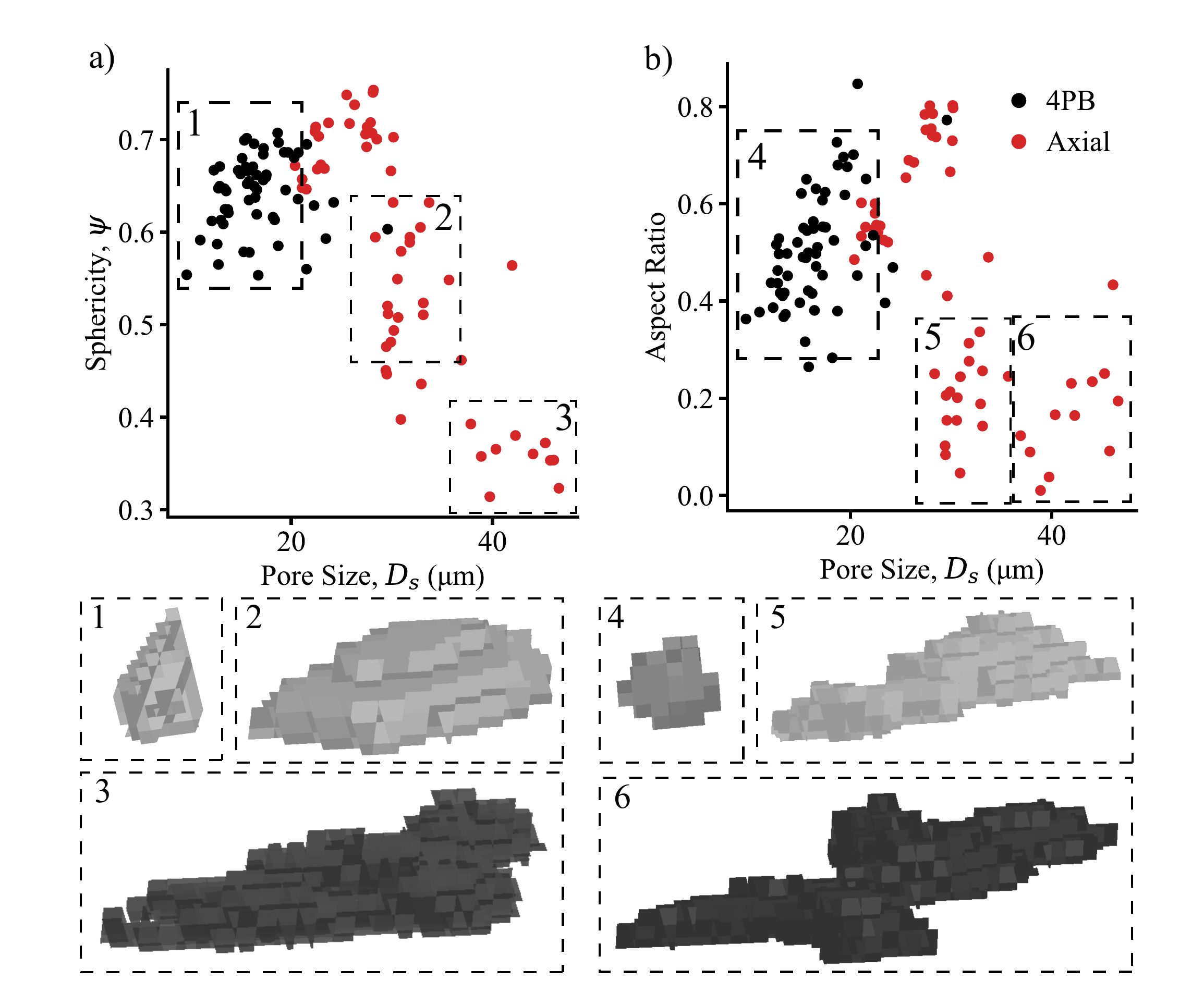}
    \caption{Scatter plot of the properties of the 30 largest pores each fatigue specimen fabricated with velocity 1300 mm/s with example pores within in each numbered region below. In the scatter plots, the specimens are differentiated by color for geometry. The porosity has 2.5 \textmu m edge length for each cubic voxel.}
    \label{fig:pores}
\end{figure}

Figure \ref{fig:pores} shows the geometric properties for pores above threshold in all fatigue specimens fabricated at 1300 mm/s, where these specimens had an unexpectedly large difference in the largest pore size distribution between the axial and 4PB geometries. These process parameters are within the expected process window for Ti-6Al-4V \cite{REDDY2024108428, Gordon2020}. From the plots in Fig. \ref{fig:pores}, the 4PB specimens contain smaller pores with higher sphericities in a) and higher aspect ratios in b). The largest pores in the 4PB specimens were highly spherical gas porosity, which is depicted in the pores from regions 1 and 4 in Fig. \ref{fig:pores}.

In the axial fatigue specimens, the pores present in the upper tail are larger and less spherical than those in the 4PB fatigue specimens. This is depicted in the pores contained in regions 2, 3, 5, and 6 where large, elongated porosity is observed. Thus, the largest pores in the axial fatigue specimens are generated from LoF, where in the 4PB fatigue specimens, no pores indicative of LoF were observed.

At the same process parameters the axial fatigue specimens contain LoF porosity, while the 4PB fatigue specimens do not. So, the differences in the largest pore size distribution between geometries in the process window are significant. This suggests that slightly smaller witness specimens at the same process parameters may not necessarily exhibit a porosity equivalence for fatigue applications. Specifically, in this case, the 4PB specimens did not contain the LoF pores that were present in the axial fatigue specimens. By not capturing the largest pore-generating phenomena, it would be impossible to predict the largest pore sizes in larger parts in the process window from smaller witness coupons. Thus, the large differences found using the EVS with uncertainty framework correspond to observable differences in porosity within the process window.

\subsection{Application to Fatigue Prediction}

\begin{figure}[H]
    \centering
    \includegraphics[width=0.9\textwidth]{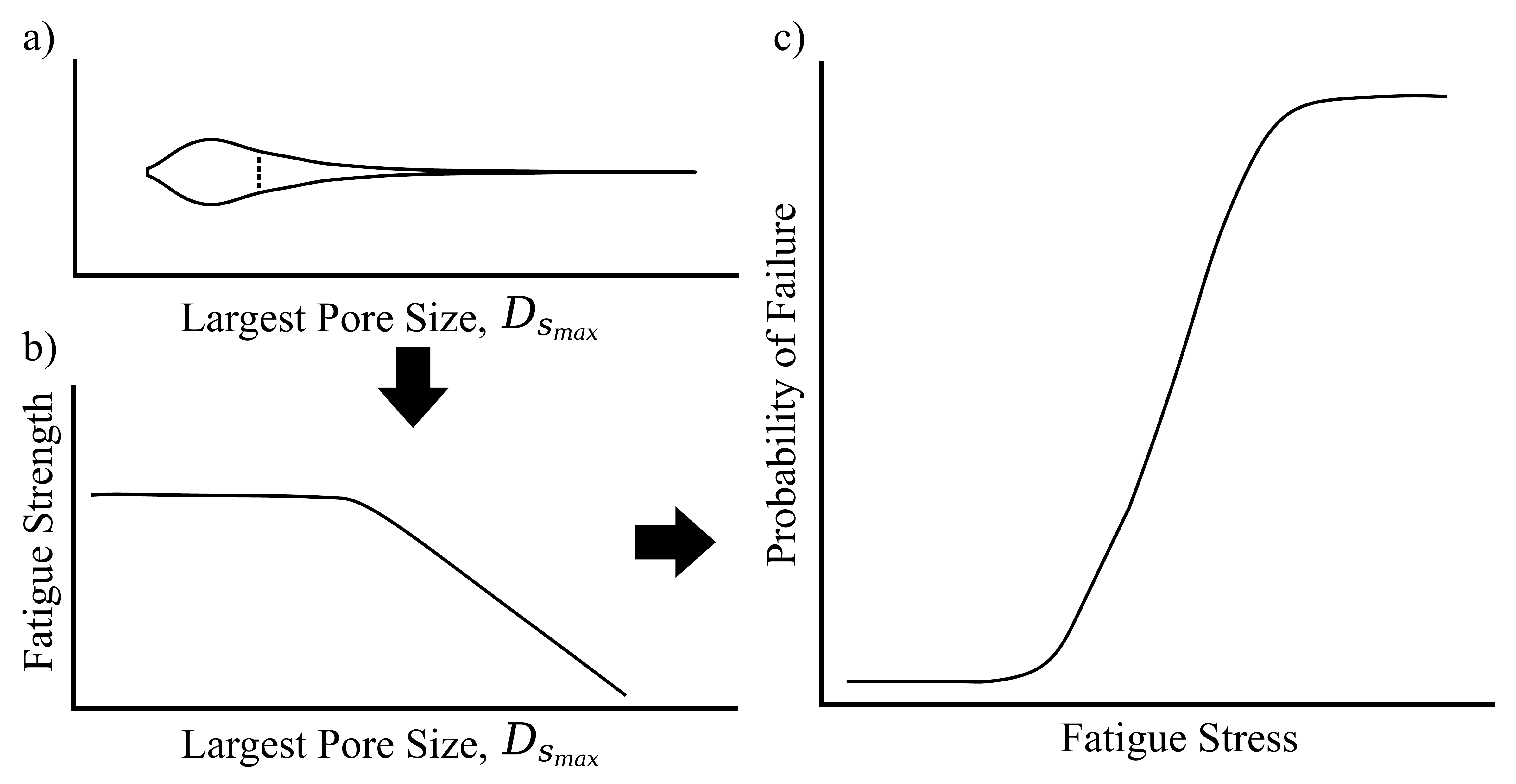}
    \caption{Diagram showing a workflow to compute scatter in fatigue properties. a) depicts the largest pore size distribution as a violin plot with the mean denoted by a dashed line. b) shows the example shape for a Kitigawa-Takahashi plot which shows the fatigue strength as a function of critical pore size.  c) contains an example probability plot showing how the distribution in a) can be propagated through the Kitigawa-Takahashi plot in b).}
    \label{fig:fatigue}
\end{figure}

By incorporating the largest pore size distributions obtained in this work into a Kitagawa-Takahashi model \cite{kitagawa1976applicability}, the scatter in fatigue strength can be directly computed. A workflow for this is demonstrated in Fig. \ref{fig:fatigue}. The ability to obtain scatter in fatigue properties in an uncertainty-driven approach is a novel application of this work. Although scatter in fatigue life has been reported in our previous work relating defects to fatigue properties \cite{REDDY2024108428}, this method can be used to capture scatter due to uncertainty in the upper tail and the resulting largest pore sizes.

\section{Conclusion}

In this work, we incorporate uncertainty quantification into EVS to estimate the largest pore size distribution in a volume of interest. To accomplish this, we captured the uncertainty in the number of pores and the shape and scale parameters of the GPD. This framework was applied to a dataset consisting of porosity from Ti-6Al-4V fatigue specimens fabricated via PBF-LB with two geometries, a 4PB specimen with rectangular prism geometry and an axial specimen with cylindrical geometry. The axial geometry had a slightly larger cross-sectional area with an as-fabricated diameter of 7 mm as opposed to 6 mm side length for the 4PB geometry. Both were fabricated under the same process conditions and X-\textmu CT was performed in the gauge section. The results showed that the EVS framework was able to estimate the largest pore size distribution in 4PB specimens, but this distribution did not always capture the largest pore sizes observed in the axial specimens fabricated with the same process parameters. Specifically, at one set of process parameters, the axial specimens contained significantly more LoF pores than the 4PB specimens, which accounted for the observed differences. Thus, the framework shown in this work is capable of determining differences in the largest pore size distribution that are relevant for fatigue-critical applications.

In addition, this work demonstrates the analysis required to establish a part-to-coupon equivalence for fatigue, which is critical to the development of a certifiable process. By incorporating the pore distributions found using this framework into existing fatigue models, the scatter in fatigue properties can be computed allowing for probabilistic prediction of fatigue properties, which an ongoing effort from the authors. The methods developed are material and process agnostic, which allows for application to other AM processes or defects of interest.

\section*{Acknowledgements}
This work was supported by the National Aeronautics and Space Administration (NASA) University Leadership Initiative (ULI) program under grant 80NSSC19M0123 and the Space Technology Research Institute (STRI) program under grant 80NSSC23K1342. 

The authors acknowledge the use of the Materials Characterization Facility at Carnegie Mellon University supported by Grant MCF-677785.

The authors acknowledge Christian Gobert and Jack Beuth for the design of experiments and fabrication. In addition, the authors acknowledge Evan Adcock and Anthony Rollett for handling post-processing and sample logistics. The authors also acknowledge Anthony Rollett for his input on connections to fatigue life and Bryan Webler for providing additional background resources.

The X-\textmu CT machine was funded by the Army Research Laboratory Cooperative Agreement Number W911NF-20-2-0175. The views and conclusions contained in this document are those of the authors and should not be interpreted as representing the official policies, either expressed or implied, of the Army Research Laboratory or the US Government. The US Government is authorized to reproduce and distribute reprints for government purposes notwithstanding any copyright notation herein. 

\section*{CRediT authorship contribution statement}

\noindent \textbf{Justin P. Miner: } Conceptualization, Methodology, Software, Validation, Formal Analysis, Investigation; Data Curation, Visualization, Writing - original draft. \textbf{Sneha Prabha Narra: } Conceptualization, Methodology, Resources, Writing - Review \& Editing, Supervision, Project Administration, Funding acquisition.

\newpage
\bibliographystyle{elsarticle-num}
\bibliography{refs}

\begin{thebibliography}{10}
\expandafter\ifx\csname url\endcsname\relax
  \def\url#1{\texttt{#1}}\fi
\expandafter\ifx\csname urlprefix\endcsname\relax\def\urlprefix{URL }\fi
\expandafter\ifx\csname href\endcsname\relax
  \def\href#1#2{#2} \def\path#1{#1}\fi

\bibitem{Tang2017}
M.~Tang, P.~C. Pistorius, J.~L. Beuth, {Prediction of lack-of-fusion porosity for powder bed fusion}, Additive Manufacturing 14 (2017) 39--48.
\newblock \href {https://doi.org/10.1016/j.addma.2016.12.001} {\path{doi:10.1016/j.addma.2016.12.001}}.

\bibitem{Cunningham2017}
R.~Cunningham, S.~P. Narra, C.~Montgomery, J.~Beuth, A.~D. Rollett, {Synchrotron-Based X-ray Microtomography Characterization of the Effect of Processing Variables on Porosity Formation in Laser Power-Bed Additive Manufacturing of Ti-6Al-4V}, Jom 69 (2017) 479--484.
\newblock \href {https://doi.org/10.1007/s11837-016-2234-1} {\path{doi:10.1007/s11837-016-2234-1}}.

\bibitem{Iebba2017}
M.~Iebba, A.~Astarita, D.~Mistretta, I.~Colonna, M.~Liberini, F.~Scherillo, C.~Pirozzi, R.~Borrelli, S.~Franchitti, A.~Squillace, {Influence of Powder Characteristics on Formation of Porosity in Additive Manufacturing of Ti-6Al-4V Components}, Journal of Materials Engineering and Performance 26 (2017) 4138--4147.
\newblock \href {https://doi.org/10.1007/s11665-017-2796-2} {\path{doi:10.1007/s11665-017-2796-2}}.

\bibitem{Snow2023}
Z.~Snow, L.~Scime, A.~Ziabari, B.~Fisher, V.~Paquit, {Observation of spatter-induced stochastic lack-of-fusion in laser powder bed fusion using in situ process monitoring}, Additive Manufacturing 61 (2023) 103298.
\newblock \href {https://doi.org/10.1016/j.addma.2022.103298} {\path{doi:10.1016/j.addma.2022.103298}}.

\bibitem{MATTHEWS201633}
M.~J. Matthews, G.~Guss, S.~A. Khairallah, A.~M. Rubenchik, P.~J. Depond, W.~E. King, Denudation of metal powder layers in laser powder bed fusion processes, Acta Materialia 114 (2016) 33--42.
\newblock \href {https://doi.org/10.1016/j.actamat.2016.05.017} {\path{doi:10.1016/j.actamat.2016.05.017}}.

\bibitem{BEREZ2022106737}
J.~Berez, L.~Sheridan, C.~Saldaña, Extreme variation in fatigue: Fatigue life prediction and dependence on build volume location in laser powder bed fusion of 17-4 stainless steel, International Journal of Fatigue 158 (2022) 106737.
\newblock \href {https://doi.org/10.1016/j.ijfatigue.2022.106737} {\path{doi:10.1016/j.ijfatigue.2022.106737}}.

\bibitem{Macallister2022}
N.~Macallister, T.~H. Becker, {Fatigue life estimation of additively manufactured Ti-6Al-4V: Sensitivity, scatter and defect description in Damage-tolerant models}, Acta Materialia 237 (2022) 118189.
\newblock \href {https://doi.org/10.1016/j.actamat.2022.118189} {\path{doi:10.1016/j.actamat.2022.118189}}.

\bibitem{PESSARD2021106206}
E.~Pessard, M.~Lavialle, P.~Laheurte, P.~Didier, M.~Brochu, High-cycle fatigue behavior of a laser powder bed fusion additive manufactured ti-6al-4v titanium: Effect of pores and tested volume size, International Journal of Fatigue 149 (2021) 106206.
\newblock \href {https://doi.org/10.1016/j.ijfatigue.2021.106206} {\path{doi:10.1016/j.ijfatigue.2021.106206}}.

\bibitem{LIU2022106700}
S.~Liu, S.~Shao, H.~Guo, R.~Zong, C.~Qin, X.~Fang, The microstructure and fatigue performance of inconel 718 produced by laser-based powder bed fusion and post heat treatment, International Journal of Fatigue 156 (2022) 106700.
\newblock \href {https://doi.org/10.1016/j.ijfatigue.2021.106700} {\path{doi:10.1016/j.ijfatigue.2021.106700}}.

\bibitem{BERETTA2021106407}
S.~Beretta, More than 25 years of extreme value statistics for defects: Fundamentals, historical developments, recent applications, International Journal of Fatigue 151 (2021) 106407.
\newblock \href {https://doi.org/10.1016/j.ijfatigue.2021.106407} {\path{doi:10.1016/j.ijfatigue.2021.106407}}.

\bibitem{REDDY2024108428}
T.~Reddy, A.~Ngo, J.~P. Miner, C.~Gobert, J.~L. Beuth, A.~D. Rollett, J.~J. Lewandowski, S.~P. Narra, Fatigue-based process window for laser beam powder bed fusion additive manufacturing, International Journal of Fatigue 187 (2024) 108428.
\newblock \href {https://doi.org/10.1016/j.ijfatigue.2024.108428} {\path{doi:10.1016/j.ijfatigue.2024.108428}}.

\bibitem{murakami1994inclusion}
Y.~Murakami, Inclusion rating by statistics of extreme values and its application to fatigue strength prediction and quality control of materials, Journal of Research of the National Institute of Standards and Technology 99~(4) (1994) 345.
\newblock \href {https://doi.org/10.6028/jres.099.032} {\path{doi:10.6028/jres.099.032}}.

\bibitem{TANG2019479}
M.~Tang, P.~C. Pistorius, Fatigue life prediction for alsi10mg components produced by selective laser melting, International Journal of Fatigue 125 (2019) 479--490.
\newblock \href {https://doi.org/10.1016/j.ijfatigue.2019.04.015} {\path{doi:10.1016/j.ijfatigue.2019.04.015}}.

\bibitem{BERETTA2017178}
S.~Beretta, S.~Romano, A comparison of fatigue strength sensitivity to defects for materials manufactured by am or traditional processes, International Journal of Fatigue 94 (2017) 178--191, fatigue and Fracture Behavior of Additive Manufactured Parts.
\newblock \href {https://doi.org/10.1016/j.ijfatigue.2016.06.020} {\path{doi:10.1016/j.ijfatigue.2016.06.020}}.

\bibitem{ROMANO201732}
S.~Romano, A.~Brandão, J.~Gumpinger, M.~Gschweitl, S.~Beretta, Qualification of am parts: Extreme value statistics applied to tomographic measurements, Materials \& Design 131 (2017) 32--48.
\newblock \href {https://doi.org/10.1016/j.matdes.2017.05.091} {\path{doi:10.1016/j.matdes.2017.05.091}}.

\bibitem{ROMANO2018165}
S.~Romano, A.~Brückner-Foit, A.~Brandão, J.~Gumpinger, T.~Ghidini, S.~Beretta, Fatigue properties of alsi10mg obtained by additive manufacturing: Defect-based modelling and prediction of fatigue strength, Engineering Fracture Mechanics 187 (2018) 165--189, sI: 50th Anniversary Issue.
\newblock \href {https://doi.org/10.1016/j.engfracmech.2017.11.002} {\path{doi:10.1016/j.engfracmech.2017.11.002}}.

\bibitem{MINERVA2023112392}
G.~Minerva, M.~Awd, J.~Tenkamp, F.~Walther, S.~Beretta, Machine learning-assisted extreme value statistics of anomalies in alsi10mg manufactured by l-pbf for robust fatigue strength predictions, Materials \& Design 235 (2023) 112392.
\newblock \href {https://doi.org/10.1016/j.matdes.2023.112392} {\path{doi:10.1016/j.matdes.2023.112392}}.

\bibitem{Fisher_Tippett_1928}
R.~A. Fisher, L.~H.~C. Tippett, Limiting forms of the frequency distribution of the largest or smallest member of a sample, Mathematical Proceedings of the Cambridge Philosophical Society 24~(2) (1928) 180–190.
\newblock \href {https://doi.org/10.1017/S0305004100015681} {\path{doi:10.1017/S0305004100015681}}.

\bibitem{gnedekno}
B.~Gnedenko, Sur la distribution limite du terme maximum d'une série aléatoire, Annals of Mathematics 44~(3) (1943) 423--453.
\newblock \href {https://doi.org/10.2307/1968974} {\path{doi:10.2307/1968974}}.

\bibitem{pickands}
J.~Pickands, III, {Statistical Inference Using Extreme Order Statistics}, The Annals of Statistics 3~(1) (1975) 119 -- 131.
\newblock \href {https://doi.org/10.1214/aos/1176343003} {\path{doi:10.1214/aos/1176343003}}.

\bibitem{balkemadehaan}
A.~A. Balkema, L.~de~Haan, {Residual Life Time at Great Age}, The Annals of Probability 2~(5) (1974) 792 -- 804.
\newblock \href {https://doi.org/10.1214/aop/1176996548} {\path{doi:10.1214/aop/1176996548}}.

\bibitem{SHAHABI2022112027}
M.~Shahabi, T.~Reddy, A.~D. Rollett, S.~P. Narra, A statistical approach to determine data requirements for part porosity characterization in laser powder bed fusion additive manufacturing, Materials Characterization 190 (2022) 112027.
\newblock \href {https://doi.org/10.1016/j.matchar.2022.112027} {\path{doi:10.1016/j.matchar.2022.112027}}.

\bibitem{EVSBook}
S.~Coles, An Introduction to Statistical Modeling of Extreme Values, 1st Edition, Springer Series in Statistics, Springer London, London, 2001.
\newblock \href {https://doi.org/10.1007/978-1-4471-3675-0} {\path{doi:10.1007/978-1-4471-3675-0}}.

\bibitem{ANDERSON201878}
K.~Anderson, S.~Daniewicz, Statistical analysis of the influence of defects on fatigue life using a gumbel distribution, International Journal of Fatigue 112 (2018) 78--83.
\newblock \href {https://doi.org/10.1016/j.ijfatigue.2018.03.008} {\path{doi:10.1016/j.ijfatigue.2018.03.008}}.

\bibitem{NASA6030}
NASA, \href{https://standards.nasa.gov/sites/default/files/standards/NASA/Baseline/0/2021-04-21_nasa-std-6030-approveddocx.pdf}{Additive manufacturing requirements for spaceflight systems}, Standard, National Aeronautics and Space Administration (2021).
\newline\urlprefix\url{https://standards.nasa.gov/sites/default/files/standards/NASA/Baseline/0/2021-04-21_nasa-std-6030-approveddocx.pdf}

\bibitem{DEVSINGH2021350}
D.~{Dev Singh}, T.~Mahender, A.~{Raji Reddy}, Powder bed fusion process: A brief review, Materials Today: Proceedings 46 (2021) 350--355, 2nd International Conference on Manufacturing Material Science and Engineering.
\newblock \href {https://doi.org/10.1016/j.matpr.2020.08.415} {\path{doi:10.1016/j.matpr.2020.08.415}}.

\bibitem{SINGH20203058}
R.~Singh, A.~Gupta, O.~Tripathi, S.~Srivastava, B.~Singh, A.~Awasthi, S.~Rajput, P.~Sonia, P.~Singhal, K.~K. Saxena, Powder bed fusion process in additive manufacturing: An overview, Materials Today: Proceedings 26 (2020) 3058--3070, 10th International Conference of Materials Processing and Characterization.
\newblock \href {https://doi.org/10.1016/j.matpr.2020.02.635} {\path{doi:10.1016/j.matpr.2020.02.635}}.

\bibitem{LEARY2021597}
M.~Leary, 22 - economic feasibility and cost-benefit analysis, in: I.~Yadroitsev, I.~Yadroitsava, A.~{du Plessis}, E.~MacDonald (Eds.), Fundamentals of Laser Powder Bed Fusion of Metals, Additive Manufacturing Materials and Technologies, Elsevier, 2021, pp. 597--620.
\newblock \href {https://doi.org/10.1016/B978-0-12-824090-8.00022-6} {\path{doi:10.1016/B978-0-12-824090-8.00022-6}}.

\bibitem{Park2024}
A.~M. Park, W.~G. Tilson, D.~N. Wells, C.~A. Kantzos, \href{https://ntrs.nasa.gov/citations/20240002004}{{Agency Additive Manufacturing ( AM ) Certification Support Team ( AACT ) Risk Reduction – Safe Life Category : Fracture Control Framework for Un-inspectable Fracture Critical AM Parts}}, Tech. Rep. February, National Aeronautics and Space Administration, Hampton, Virginia (2024).
\newline\urlprefix\url{https://ntrs.nasa.gov/citations/20240002004}

\bibitem{ASTM2012}
ASTM, {ASTM F3001-14 : Standard Specification for Additive Manufacturing Titanium-6 Aluminum-4 Vanadium ELI (Extra Low Interstitial) with Powder Bed Fusion} (2012).
\newblock \href {https://doi.org/10.1520/F3001-14R21} {\path{doi:10.1520/F3001-14R21}}.

\bibitem{ams28012003heat}
A.~G. Titanium, R.~M. Committee, Heat Treatment of Titanium Alloy Parts (2014).
\newblock \href {https://doi.org/10.4271/AMS2801B} {\path{doi:10.4271/AMS2801B}}.

\bibitem{ASTM2002}
ASTM, {Standard Practice for Conducting Force Controlled Constant Amplitude Axial Fatigue Tests of Metallic Materials}, Test 03 (2002) 4--8.
\newblock \href {https://doi.org/10.1520/E0466-21.2} {\path{doi:10.1520/E0466-21.2}}.

\bibitem{ObjectResearchSystemsORSInc2022}
{Comet Technologies Canada Inc.}, \href{http://www.theobjects.com/dragonfly}{{Dragonfly 2022.2}} (2022).
\newline\urlprefix\url{http://www.theobjects.com/dragonfly}

\bibitem{unetct}
V.~W.~H. Wong, M.~Ferguson, K.~H. Law, Y.-T.~T. Lee, P.~Witherell, {Segmentation of Additive Manufacturing Defects Using U-Net}, Journal of Computing and Information Science in Engineering 22~(3) (2021) 031005.
\newblock \href {https://doi.org/10.1115/1.4053078} {\path{doi:10.1115/1.4053078}}.

\bibitem{Desrosiers2024}
C.~Desrosiers, M.~Letenneur, F.~Bernier, N.~Pich{\'e}, B.~Provencher, F.~Cheriet, F.~Guibault, V.~Brailovski, Automated porosity segmentation in laser powder bed fusion part using computed tomography: a validity study, Journal of Intelligent Manufacturing (2024).
\newblock \href {https://doi.org/10.1007/s10845-023-02296-w} {\path{doi:10.1007/s10845-023-02296-w}}.

\bibitem{Poisson}
H.~Madsen, P.~F. Rasmussen, D.~Rosbjerg, Comparison of annual maximum series and partial duration series methods for modeling extreme hydrologic events: 1. at-site modeling, Water Resources Research 33~(4) (1997) 747--757.
\newblock \href {https://doi.org/10.1029/96WR03848} {\path{doi:10.1029/96WR03848}}.

\bibitem{rosbjerg1992prediction}
D.~Rosbjerg, H.~Madsen, P.~F. Rasmussen, Prediction in partial duration series with generalized pareto-distributed exceedances, Water Resources Research 28~(11) (1992) 3001--3010.
\newblock \href {https://doi.org/10.1029/92WR01750} {\path{doi:10.1029/92WR01750}}.

\bibitem{HU2023108935}
X.~Hu, G.~Fang, J.~Yang, L.~Zhao, Y.~Ge, Simplified models for uncertainty quantification of extreme events using monte carlo technique, Reliability Engineering \& System Safety 230 (2023) 108935.
\newblock \href {https://doi.org/10.1016/j.ress.2022.108935} {\path{doi:10.1016/j.ress.2022.108935}}.

\bibitem{GHOSH20101492}
S.~Ghosh, S.~Resnick, A discussion on mean excess plots, Stochastic Processes and their Applications 120~(8) (2010) 1492--1517.
\newblock \href {https://doi.org/10.1016/j.spa.2010.04.002} {\path{doi:10.1016/j.spa.2010.04.002}}.

\bibitem{BRAZAUSKAS2009424}
V.~Brazauskas, A.~Kleefeld, Robust and efficient fitting of the generalized pareto distribution with actuarial applications in view, Insurance: Mathematics and Economics 45~(3) (2009) 424--435.
\newblock \href {https://doi.org/10.1016/j.insmatheco.2009.09.002} {\path{doi:10.1016/j.insmatheco.2009.09.002}}.

\bibitem{nthorder}
B.~C. Arnold, N.~Balakrishnan, H.~N. Nagaraja, 4. Order Statistics from Some Specific Distributions, Classics in Applied Mathematics, Society for Industrial and Applied Mathematics, 2008, Ch.~4, pp. 63--106.
\newblock \href {https://doi.org/10.1137/1.9780898719062.ch4} {\path{doi:10.1137/1.9780898719062.ch4}}.

\bibitem{Huang2012}
C.~Huang, J.-G. Lin, Y.-Y. Ren, Statistical inferences for generalized pareto distribution based on interior penalty function algorithm and bootstrap methods and applications in analyzing stock data, Computational Economics 39~(2) (2012) 173--193.
\newblock \href {https://doi.org/10.1007/s10614-011-9256-0} {\path{doi:10.1007/s10614-011-9256-0}}.

\bibitem{das2016characterization}
K.~P. Das, C.~A. Sams, V.~Singh, \href{http://jsr.isrt.ac.bd/wp-content/uploads/48_50n2_4.pdf}{Characterization of the tail of river flow data by generalized pareto distribution}, Journal of Statistical Research 48~(50) (2016) 2.
\newline\urlprefix\url{http://jsr.isrt.ac.bd/wp-content/uploads/48_50n2_4.pdf}

\bibitem{evsmc}
X.~G. Larsén, J.~Mann, O.~Rathmann, H.~E. Jørgensen, Uncertainties of the 50-year wind from short time series using generalized extreme value distribution and generalized pareto distribution, Wind Energy 18~(1) (2015) 59--74.
\newblock \href {https://doi.org/10.1002/we.1683} {\path{doi:10.1002/we.1683}}.

\bibitem{cupy_learningsys2017}
R.~Okuta, Y.~Unno, D.~Nishino, S.~Hido, C.~Loomis, \href{http://learningsys.org/nips17/ass ets/papers/paper\_16.pdf}{Cupy: A numpy-compatible library for nvidia gpu calculations}, in: Proceedings of Workshop on Machine Learning Systems (LearningSys) in The Thirty-first Annual Conference on Neural Information Processing Systems (NIPS), 2017, p.~1.
\newline\urlprefix\url{http://learningsys.org/nips17/ass ets/papers/paper\_16.pdf}

\bibitem{smallinterval}
M.~B. Nejc~Bezak, M.~Šraj, Comparison between the peaks-over-threshold method and the annual maximum method for flood frequency analysis, Hydrological Sciences Journal 59~(5) (2014) 959--977.
\newblock \href {https://doi.org/10.1080/02626667.2013.831174} {\path{doi:10.1080/02626667.2013.831174}}.

\bibitem{MORAN2021102333}
T.~Moran, D.~Warner, A.~Soltani-Tehrani, N.~Shamsaei, N.~Phan, Spatial inhomogeneity of build defects across the build plate in laser powder bed fusion, Additive Manufacturing 47 (2021) 102333.
\newblock \href {https://doi.org/10.1016/j.addma.2021.102333} {\path{doi:10.1016/j.addma.2021.102333}}.

\bibitem{Gordon2020}
J.~V. Gordon, S.~P. Narra, R.~W. Cunningham, H.~Liu, H.~Chen, R.~M. Suter, J.~L. Beuth, A.~D. Rollett, {Defect structure process maps for laser powder bed fusion additive manufacturing}, Additive Manufacturing 36 (2020) 101552.
\newblock \href {https://doi.org/10.1016/j.addma.2020.101552} {\path{doi:10.1016/j.addma.2020.101552}}.

\bibitem{kitagawa1976applicability}
H.~Kitagawa, S.~Takahashi, Applicability of fracture mechanics to very small cracks or the cracks in the early stage, in: Proc. 2nd int. conf. on mechanical behaviour of materials, 1976, pp. 627--631.

\end{thebibliography}

\newpage
\appendix

\section{Understanding the Volumetric Scaling Effect on Largest Pore Size}
\label{sec:volumeeffect}

\begin{figure}[H]
    \centering
    \includegraphics[width=0.85\textwidth]{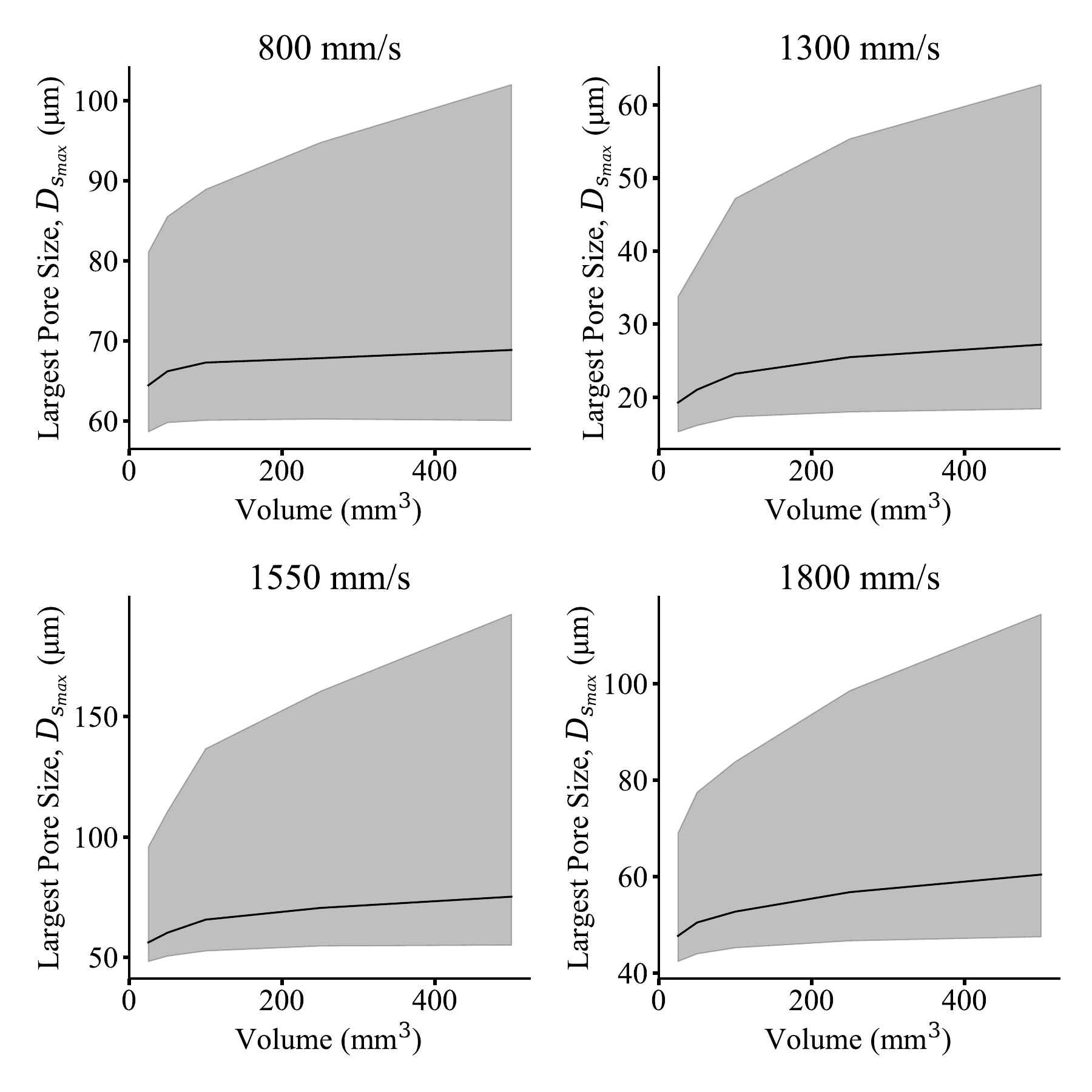}
    \caption{Plot showing how the largest pore size estimation changes with volume. The 95\% confidence interval is shown with the gray ribbon. It can be seen that the mean size plateaus with increasing volume, where the upper bound continues to increase.}
    \label{fig:volumescale}
\end{figure}

Figure \ref{fig:volumescale} shows how the estimated largest pore size changes with respect to increasing volume of interest. It can be observed that the mean line plateaus with increasing volume among nearly all specimens, however, the upper bound of the confidence interval or the 97.5\textsuperscript{th} percentile pore size at a given volume continues to increase. This highlights that as the volume of interest increases, uncertainty in the largest pore size will increase, and larger pores could be expected, which aligns with what is expected.

\end{document}